\documentclass[aps,prb,superscriptaddress,twocolumn,longbibliography]{revtex4-1}
\usepackage{amsmath}
\usepackage{amssymb}
\usepackage{graphicx}
\usepackage[usenames,dvipsnames]{xcolor}
\usepackage{tikz}
\usepackage{pgffor}
\usepackage{verbatim}
\usepackage{float}
\usepackage{mathrsfs}
\usepackage[colorlinks, breaklinks=true,linkcolor=red, citecolor=blue, linktocpage=true]{hyperref}
\usepackage{cleveref}
\usepackage{tensor}
\usepackage{eufrak}

\usepackage{color}

\newcommand*{\citen}[1]{%
  \begingroup
    \romannumeral-`\x 
    \setcitestyle{numbers}%
    \cite{#1}%
  \endgroup   
}


\begin{document}

\title{Analogue Stochastic Gravity in Strongly-Interacting Bose-Einstein Condensates} 
\author{Aydin Cem Keser}
 \affiliation{Condensed Matter Theory Center and Joint Quantum Institute, University of Maryland, College Park, MD 20742, USA}
 \affiliation{School of Physics and Australian Research Council Centre of
 	Excellence in Low-Energy Electronics Technologies,
 	UNSW  Node,  The University  of New  South  Wales,  Sydney  2052,  Australia}
\author{Victor Galitski}
 \affiliation{Condensed Matter Theory Center and Joint Quantum Institute, University of Maryland, College Park, MD 20742, USA}
 \affiliation{School of Physics, Monash University, Melbourne, Victoria 3800, Australia.}

\date{\today}

\begin{abstract}
Collective modes propagating in a moving superfluid are known to satisfy wave equations in a curved space time, with a metric determined by the underlying superflow. We use the Keldysh technique in a curved space-time to develop a quantum geometric  theory of fluctuations in superfluid hydrodynamics. This theory relies on a ``quantized'' generalization of the two-fluid description of Landau and Khalatnikov, where the superfluid component is viewed as a quasi-classical field coupled to a normal component -- the collective modes/phonons representing a quantum bath. This relates the problem in the hydrodynamic limit to the ``quantum friction'' problem of Caldeira-Leggett type. By integrating out the phonons, we derive stochastic Langevin equations describing a coupling between the superfluid component and phonons. These equations have the form of Euler equations with additional source terms expressed through a fluctuating stress-energy tensor of phonons.   Conceptually, this result is similar to stochastic Einstein equations that arise in the theory of stochastic gravity. We formulate the fluctuation-dissipation theorem in this geometric language and  discuss possible physical consequences of this theory.
\end{abstract}
\maketitle

\section {Introduction}

The idea that a curved space-time is an emergent structure has a long history~\cite{sakharov, Visser_sakharov} and has been discussed in various physical contexts~\cite{analogue_gravity_review} from classical fluid mechanics~\cite{Unruh,Stone} and crystals with defects~\cite{Kleinert_defect} to quantum entanglement.~\cite{entanglement} While in the context of fundamental gravity, the emergent scenario remains speculative at this stage,  there has been a number of concrete realizations of various aspects of general relativity in ``analogue gravity'' models, where a non-trivial curved space-time metric arises naturally in the description of collective modes relative to a background solution of the field equations.~\cite{analogue_gravity_review,analogue_spacetimes} A prominent example of such analogue theory is a strongly-correlated superfluid,~\cite{Volovik_book} where the phonon modes propagating relative to a (generally inhomogeneous and non-stationary) superflow satisfy a wave-equation  in an effective curved space-time
\begin {equation}
\label{cov_wave}
\partial_\mu \left( \sqrt{-g} g^{\mu\nu} \partial_\nu \phi \right) = 0,
\end {equation}
where $\phi(\vec{r},t)$ is the phonon field -- a small deviation from a ``mean-field'' configuration, $g = \text{det}{g_{\mu\nu}}$ is the determinant and  $g^{\mu\nu}$ is the matrix  inverse of the metric 
\begin {equation}
\label{metric}
g_{\mu\nu} = \frac{\rho}{c}
\begin{pmatrix}
c^2 - v^2 & \vec{v}^T\\
\vec{v} & -\mathbf{I}_{3\times 3}
\end{pmatrix}
\end {equation}
which is determined by the underlying superflow ($\vec{v}$ and $c$ are the superfluid velocity and the speed of sound, $\rho$ is the density of the fluid including the excitations). Many exciting general-relativistic effects immediately follow from this observation, including the formation of sonic horizons and black hole-type physics,\cite{analogue_spacetimes} analogue Hawking radiation,~\cite{Hawking_BEC} proposed by Unruh~\cite{Unruh} and recently  reported by Steinhauer to have been observed  in cold-atom Bose-Einstein condensates,~\cite{steinhauer} and a unifying principle for cosmology and high energy physics discussed extensively by Volovik.~\cite{Volovik_book,Volovik_cosmology}

This geometric theory of excitations in a superfluid, that this paper focuses on and develops further, is an alternative formulation of the phenomenological Landau-Khalatnikov two-fluid theory of superfluidity,~\cite{LLfluid, Khalatnikov} which has been originally developed as a macroscopic description of superfluid Helium. As the name suggests, this theory separates the fluid flow into two components -- one being the zero entropy, zero viscosity superflow and the other being the entropy-carrying, dissipative normal flow.  The two-fluid theory relies on the conservation laws for mass, energy, and momentum in a Galilean-invariant continuum made up of these two components. In addition to being an accurate macroscopic description of superfluid helium, the two-fluid theory can be viewed as the first historical example of a long wavelength hydrodynamic limit of a strongly interacting quantum field theory. The low energy effective field theory paradigm offers a number of powerful techniques to analyze strongly interacting field theories and the hydrodynamic limit of high energy theories (e.g., the AdS/CFT and string theory).~\cite{Kadanoff, M-theory, brane_viscosity, Ads/CFT}

The main idea of this work relies on a conceptual analogy between the quantum generalization of the Landau-Khalatnikov two-fluid description and the Caldeira-Leggett-type theories of ``quantum friction'', where a closed system is separated into two components -- a quantum ``particle'' and a bath to which it is coupled.~\cite{weiss, Altland_book} Integrating out the bath leads to classical equations of motion for the particle, which necessarily feature a friction force and a stochastic Langevin force, connected to each other via a fluctuation-dissipation theorem. For a strongly-correlated BEC, this analogy associates the superfluid order parameter field with the Caldeira-Leggett ``particle'' and the Bogoliubov excitations with the bath. The question we ask here is what is the corresponding Langevin equations of motion that arise? We develop and use a combination of the aforementioned geometric theory of excitations and Keldysh field-theoretical methods in a curved space time to answer this question. The main result is the following stochastic equations of motion 
\begin {subequations}
\label{euler}
\begin {align}
\label{cont}
&\partial_t \rho + \nabla \cdot(\rho \vec{v}) = \frac{1}{2} \nabla \cdot \left[\sqrt{-g} \left(  \left\langle \hat{T}_{\mu\nu} \right\rangle + \xi_{\mu \nu} \right)   \frac{\delta g^{\mu\nu}}{\delta \vec{v}}\right],\\
\label{Bernouilli}
&\partial_t \theta + \frac{1}{2} v^2 + \mu(\rho)  = \frac{\sqrt{-g}}{2} \left(  \left\langle \hat{T}_{\mu\nu} \right\rangle + \xi_{\mu \nu} \right)  \frac{\delta g^{\mu\nu}}{\delta \rho},
\end {align}
\end {subequations}
where $\rho$ is the density of the fluid, $\theta$ is the superfluid flow potential, $\vec{v} = \nabla \theta$ is the irrotational flow fluid of the superfluid component, $\mu(\rho)$ is the chemical potential of the fluid, $g^{\mu\nu}$ is the matrix inverse of the metric tensor~\eqref{metric}, $\sqrt{-g}=|\text{det}{(g_{\mu\nu}})|^{1/2}$ is the space-time volume measure, $\hat{T}_{\mu\nu}(x)$ is the stress-energy tensor operator of the phonons (here $x$ is a short-hand for the $(3+1)$-space-time variable and the Einstein summation convention is in use) and $ \xi_{\mu \nu}(x)$ is a stochastic tensor field, describing its fluctuations around the average $\langle \hat{T}_{\mu\nu}(x) \rangle$. That is, the statistics of the Gaussian noise  with zero average is determined by the correlator
$$
\left\langle \xi_{\mu \nu}(x)\xi_{\mu' \nu'}(x') \right\rangle = \frac{1}{2}\left\langle \left\{ \hat{t}_{\mu \nu} (x) ,  \hat{t}_{\mu' \nu'}(x')  \right\} \right\rangle,
$$
where $\hat{t}_{\mu \nu} (x) = \hat{T}_{\mu\nu}(x) -\langle \hat{T}_{\mu\nu}(x) \rangle$ and $\left\{ \cdot, \cdot \right\}$ is the anti-commutator. The averages here are calculated relative to a deterministic background. What these equations actually describe are fluctuations in the superfluid, e.g. they yield statistics of density and velocity fluctuations, which in turn determine  a stochastic metric. In this sense, there is a strong similarity to the stochastic Einstein equations discussed n the context of stochastic gravity.~\cite{Hu_Verdaguer_primer,Martin_Verdaguer_stochastic_gravity} 

While these stochastic Einstein equations are interesting in and by themselves, their derivation presents a technical challenge (a non-trivial generalization of the non-equilibrium Keldysh techniques for a curved space-time is required)~\cite{Calzetta_Hu_CPT} and gives rise to a number of additional interesting results along the way, as discussed below. Our paper is structured as follows:

In Sec.~\ref{sec:vacuum}, we discuss the applicability of the metric description of a superfluid by analyzing the relevant length and energy scales. In analogy with cosmology, the geometric description breaks down at an effective ``Planck energy,'' where both the linear dispersion of phonons and the hydrodynamic description break down. 

In Sec.~\ref{sec:matter}, we use the background field formalism to write down the Keldysh quantum field theory of quasiparticles. We emphasize that the Keldysh description is necessary for taking the dynamical fluctuations of the phonon field into account.

In Sec.~\ref{sec:two-fluid}, we derive the analogue Einstein equation that governs the background and the excitations -- ``matter field.'' We establish  equivalence of the analogue Einstein equation and the covariant conservation law for the phonons to the  two-fluid conservation laws of Landau and Khalatnikov.  We prove the equivalence of the two descriptions by reducing the covariant conservation law  down to the Noether current of the two-fluid system by using the equations of motion. In Appendix~\ref{app:conservation}, we provide the technical details of this derivation.

 In Sec.~\ref{sec:stochastic}, we take the analogy between the superfluid system and general relativity further to the domain of stochastic fluctuations. We write the response and dissipation kernels in the covariant language, and give the details of this in Appendix~\ref{app:Kernels}. In global thermal equilibrium, we discuss the notion of temperature on curved space-time. We prove the fluctuation dissipation relation for a metric with globally time-like Killing vectors, that is for a flow that can brought to a stationary form after a  Galilean transformation.

Finally in Sec.~\ref{minkowski}, we linearize the stochastic analogue Einstein equation  and obtain a Langevin-type equation for the stochastic corrections to the background. We show that the symmetries of the flow determine structure of the Langevin equation, by considering the Minkowski case.

Throughout the paper, we will use the Einstein summation convention for the indices, unless otherwise stated. The space-time indices are in small case Greek letters while the space indices are in small case Latin letters. We use the sign convention $(+ -  - -)$.  In addition, the fluid dynamics equations are written in terms of Cartesian tensors, where the distinction between covariant and contravariant tensors is not important.

\section{The model and energy scales}
\label{sec:vacuum}

In this section, we review the energy scales involved in the analysis of an interacting system of bosons and its excitations.
The analogue``general relativistic'' description is an approximation to the exact theory and its applicability  is controlled by our ability, or lack thereof, to neglect a quantum pressure term discussed below. The main conclusion of this section is that the stronger the repulsive interactions between bosons composing the superfluid, the less important the quantum pressure term and correspondingly the wider the domain of applicability of the  general-relativistic approximation (in the sense of a range of energies and length-scales where the  description applies). We discuss these ``Planck'' energy and length-scales below.   

Our starting point is just the standard Lagrangian  of interacting bosons 
\begin {equation}
\label{boson}
\mathscr{L}[\Phi,\Phi^*] = \Phi^* i\hbar \partial_t \Phi - \frac{\hbar^2}{2 m }|\nabla \Phi|^2 -\varepsilon\left(|\Phi|^2\right), 
\end {equation}
where $\Phi(\vec{r},t) \equiv \Phi(x)$ is the boson field, $m$ is the mass of a boson, and the energy $\varepsilon(|\Phi|^2)$ describes an external potential and density-density repulsion between the bosons. Though, at this stage we do not specify the external potential and  interaction potential between bosons, for $\varepsilon = \frac{g}{2} |\Phi|^4 + V |\Phi|^2$, the saddle point of this Lagrangian, satisfies the Gross-Pitaevskii or non-linear Schr{\"o}dinger equation:
\begin {equation}
\label{GPE}
i\hbar \partial_t \Phi = -\frac{\hbar^2}{2 m}\nabla^2 \Phi + V(\vec{r},t) \Phi+ g |\Phi|^2 \Phi.
\end {equation} 
The first step in deriving the hydrodynamic theory is the Madelung transformation of the boson field,~\cite{Madelung1927} which is a change of variables to polar coordinates in each point of space-time:
\begin {equation}
\Phi(x) = \sqrt{\frac{ \rho(x)}{m}} e^{i m \theta(x)/\hbar}.
\end {equation}
The  new variables are the density, $\rho$, and the phase $\theta$, which in effect is a ``flow potential'' for the superfluid, that gives rise to the irrotational flow velocity field
$$
\vec{v} = \nabla {\theta}.
$$  
In terms of these variables, the Lagrangian~\eqref{boson} takes the form
\begin {equation}
-\mathscr{L} = \rho \partial_t \theta + \frac{1}{2}\rho \vec{v}^2 + \varepsilon(\rho)  + \!\!\!\!\! \underbrace{\frac{1}{8} \left(\frac{\hbar}{m} \frac{\nabla \rho}{\sqrt{\rho}}\right)^2}_{\small \mbox{``quantum pressure''}}.
\label{Lagrange}
\end {equation}

Classically, the long-wavelength description of this system in local thermodynamic equilibrium is fluid dynamics. This description can be extended to the quantum regime, where a macroscopic order exists as in the case of a condensate. This macroscopic order  is a non-trivial vacuum that solves the mean field fluid equation Eq.~\eqref{GPE_hydro}, which is just the Gross-Pitaevskii equation Eq.~\eqref{GPE} in the Madelung parametrization:
\begin {subequations}
\label{GPE_hydro}
\begin {align}
\label{continuity}
\partial_t \rho + \nabla\cdot(\rho  \vec{v}) &= 0,\\
\label{Bernouilli_madelung}
\partial_t \theta + \frac{1}{2}\vec{v}^2 + \frac{\partial\varepsilon}{\partial\rho} &= \frac{1}{2} \frac{\hbar^2}{m^2} \frac{1}{\sqrt{\rho}} \nabla^2 \sqrt{\rho}.
\end {align}
\end {subequations}
These are the Euler equations for an ideal, zero entropy fluid, with an additional energy per unit mass appears in the right-hand side of Eq.~\eqref{Bernouilli_madelung}  due to the  quantum potential term  $\sim (\nabla \sqrt{\rho})^2$ in Eq.~\eqref{Lagrange}. Together with the continuity equation Eq.~\eqref{continuity}, the gradient of ~\eqref{Bernouilli_madelung}, when multiplied by the density $\rho$,  produces a momentum balance equation. In this equation, the quantum potential leads to a pressure gradient, hence  this potential is called ``quantum pressure'' in Eq.~\eqref{Lagrange}.

The excitations over this non-trivial vacuum state are the Bogoluibov quasiparticles. These quasiparticles can be thought of a quantum field propagating on top of the ground state manifold. When treated semiclassically, the quasiparticles constitute sources to the hydrodynamic equations~\eqref{GPE_hydro}, in the spirit of the two-fluid hydrodynamics. Moreover, the entropy-carrying quasiparticle quantum field acts as a bath on the background system. In addition to momentum and energy flux and stresses, the quantum bath creates a noise to the evolution of the background, leading to a stochastic Langevin component.
   
We now show, by analyzing the appropriate scales, that when the repulsive interactions between the bosons are strong, the quantum pressure term is suppressed. Such a fluid is the basis of the gravitational analogy, as it simulates the space-time on which the matter field -- that is, the phonon field -- propagates.~\cite{Volovik_book, Volovik_cosmology}

If we momentarily ignore the quantum pressure proportional to $\nabla \sqrt{\rho}$, we get the Lagrangian density of a vortex free perfect fluid with flow velocity $\vec{v} = \nabla \theta$. The excitations of this system are obtained by linearizing the equations of motion and are the sound waves with speed $c^2 = \rho d^2 \varepsilon / d \rho^2$ at equilibrium density. When quantized, the excitations have the linear spectrum $E_p = \hbar c k$. Therefore, the dimensional parameters characterizing  the vortex free quantum hydrodynamics are the Planck constant, $\hbar$, the equilibrium density $\rho_e$, and the equilibrium speed of sound, $c_e$. The relevant energy scale of this theory is $E_Q = (\hbar^3 c_e^5 \rho_e)^{1/4}$. Since the energy density is $\sim \rho_e c_e^2$, the characteristic length scale  is $d_Q = (\rho_e c_e / \hbar)^{-1/4}$. Therefore, the mass scale is $\rho_e d_Q^3 = M_Q = (\rho_e \hbar^3 / c_e^3)^{1/4}$ and the time-scale is $t_Q = d_Q/c_e =(\rho_e c_e^5 / \hbar)^{-1/4}$.

With the addition of the quantum pressure term, 
the spectrum of Bogoliubov phonons receives a correction as $E_p = \hbar c_e k \sqrt{1 + \hbar^2 k^2 /(4 m^2 c_e^2)}$. Therefore the linear spectrum of phonons breaks down at a length scale of $\xi = \hbar/(m c_e)$, this is the coherence length of the condensate. At this scale, the phonon energy is of order $E_{Lorentz} = mc^2$, which can be dubbed the Lorentz violation energy (i.e., where the phonon spectrum deviates from the linear dispersion). Note that, Lorentz violations do not necessarily occur at exactly the inter atomic length scale $d$ and thus $E_{Lorentz}$ is generally distinct  from the ``Planck'' energy scale $E_{Planck} = \hbar c_e /d$  -- that is the energy required to resolve the individual atoms separated by a distance $d$ (at such length-scales the hydrodynamic description becomes meaningless). Indeed, the ratio of the coherence length to the inter atomic distance is determined by the strength of the atomic interactions. Defining $g_0  = \hbar^2 d /m^3$, and setting $g \rho_e^2 = \varepsilon (\rho_e)  = \rho_e c_e^2$, we get $(\xi/d)^2 = g_0/g$. In summary the relationships between different scales can be written in terms of the normalized strength of interactions $g_0/g$ as:

\begin {equation}
\frac{E_{Planck}}{E_{Lorentz}} = \frac{\xi}{d} = \sqrt{\frac{g_0}{g}} = \left(\frac{M_Q}{m}\right)^{4/3}.
\label{scales}
\end {equation} 

This ratio determines the relative importance of the quantum pressure term compared to the interaction energy. The corresponding ratio is (below, $L$ is a length-scale on which the density changes):
$$
\frac{ \left( \frac{\hbar}{m} \nabla \sqrt{\rho}\right)^2}{\varepsilon({\rho})} \sim
\frac{\hbar^2}{g m^2 \rho L^2 } \sim \frac{\xi^2}{L^2} \lesssim  \frac{\xi^2}{d^2} = \frac{g_0}{g}.
$$

In the weak interaction limit, as in a dilute Bose gas, the coherence length $\xi$ and the quantum mass scale $M_Q$ are large compared to microscopic counterparts $d$ and $m$. This signals that  the system behaves like a macroscopic quantum object, hence the condensate fraction is closed to unity. In this regime, Lorentz violation occurs much before the  interatomic scales are reached and the excitations are Bogoluibov quasiparticles with the non-linear spectrum. In the strong interaction regime, as in Helium-II or a strongly-interacting BEC, the condensate fraction is small, and the Lagrangian (\ref{Lagrange}) without the quantum pressure term describes the superfluid, as it produces correct equations for a zero entropy dissipationless superfluid.  In this regime, the quantum pressure is negligible  down to the $\xi \sim d$ scale. This means that the collective excitations are sound waves all the way up to the effective Planck energy. Therefore, geometric theory of analogue gravity for the covariant sound waves, that we will summarize in the next section, applies as long as we stay in the hydrodynamic regime, that is to say at length scales larger than the inter atomic distance.

\section{Background field formalism on the Keldysh contour for Bogoliubov quasiparticles}
\label{sec:matter}

Here, we outline a procedure  to extract a field theory of the excitations starting from Eq.~\eqref{Lagrange}. We  show that an effective curvature and covariance emerge when the amplitude modes of the excitations are integrated out. Finally, we obtain an effective action for the superfluid system and the phonon bath. 
Unless noted otherwise, we use the units, where
\begin {equation}
\label{units}
\hbar = c_e = k_B = 1.
\end {equation}

We employ the closed-time path integral or the Keldysh functional  integral~\cite{Kadanoff_Baym,Keldysh,Konstantinov,Schwinger} and the background field 
formalism~\cite{Peskin, Abbott} to separate the superfluid background and the quantum field theory of excitations. Note that the procedure, outlined below in Eqs.~\eqref{rho(t)} -- \eqref{L2_def} is completely general and does not rely on a particular form of the initial action, but we will apply it specifically to the superfluid Lagrangian~\eqref{Lagrange}.

\begin{figure}[h]
\begin{center}$
\begin{array}{cc}
\includegraphics[width=0.45\textwidth]{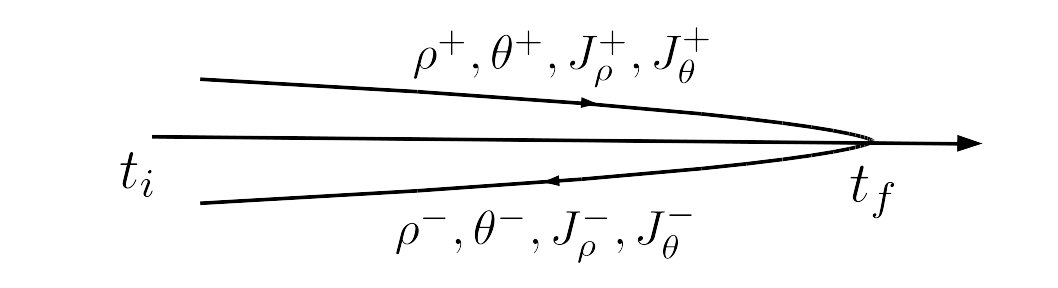}
\end{array}$
\end{center}
\caption{The Keldysh contour and its forward/backward branches labeled by the path index $s=\pm$: Each degree of freedom and the associated sources are defined twice, one on each branch. The values are matched at $t_f$, for example $\rho_f^+$ = $\rho_f^-$.}
\label{Fig1}
\end{figure}
Generally, given an initial density matrix $\boldsymbol{\hat{\varrho}}(t_i)$  and   the evolution operator $\hat{\mathscr{U}}_{t',t}$ that takes the system from $t$ to $t'$,  the density matrix at any point in time $t$ is 
\begin {equation}
\label{rho(t)}
\boldsymbol{\hat{\varrho}}(t) = \mathscr{\hat{U}}_{t,t_i}\boldsymbol{\hat{\varrho}}(t_i)\mathscr{\hat{U}}_{t_i,t} .
\end {equation}
Suppose $\mathscr{\hat{O}}$ is an observable. In  the Schr{\"o}dinger picture, the expectation value of this operator at time $t$ is
\begin {equation}
\label{expectation}
\left\langle \mathscr{hat{O}} (t) \right\rangle = \text{Tr}\, \left[ \mathscr{O}\boldsymbol{\hat{\varrho}}(t) \right].
\end {equation}

In the Keldysh formalism, the time contour is made up of two branches that start at an early time $t_i$ and meet at $t_f$ as shown in Fig~\ref{Fig1}. Each degree of freedom in the system is defined twice, one on the forward and one on the backward branch of the time contour labeled by $\pm$ respectively. The values of all variables are matched at the final point $t_f$ where the branches meet.   The observable can be coupled to the system via the source currents $J^\pm (\vec{r},t) $ that are added to the Hamiltonian $\hat{H} \to \hat{H} + J^\pm \mathscr{\hat{O}}$. The forward/backward evolution operators associated with the modified Hamiltonian are denoted as $\mathscr{\hat{U}} (J^\pm)$. Then the forward/backward expectation values of the observable can be obtained from the following generating function
\begin {equation}
\label{generating}
\mathscr{Z} [{J}^{\pm}]  = \text{Tr}\, \left[\mathscr{\hat{U}}_{t_i,t_f}({J}^-)\mathscr{\hat{U}}_{t_f,t_i}({J}^+) \boldsymbol{\hat{\varrho}}(t_i) \right],
\end {equation}
by differentiating it with respect to the forward/backward source currents $J^\pm$
\begin {equation}
\label{pm_observable}
\left\langle \mathscr{hat{O}}^\pm (t) \right\rangle  = \pm  i \frac{\delta }{\delta J^\pm (t)}\mathscr{Z} [{J}^{\pm}] \bigg\vert_{J^\pm = 0}.
\end {equation}

Note that, if we set ${J}^+ = {J}^-$ from the outset,  the forward/backward evolution operators cancel due to unitarity and 
\begin {equation}
\label{keldysh_unitary}
\mathscr{Z} [\vec{J}^+ = \vec{J}^- ] = \text{Tr}\, \boldsymbol{\hat{\varrho}}(t_i)  = 1.
\end {equation}
This means taking the logarithm of the generating function as in ordinary field theory, is redundant in Keldysh theory.

The generating  function Eq.~\eqref{generating} for the boson system in Eq.~\eqref{Lagrange} can be written as a closed time path integral as~\cite{Kamenev_book,Altland_book}:
\begin {multline}
\label{fullZ}
\!\!\!\! \mathscr{Z} [J_\rho^{\pm},J_\theta^{\pm}] = \int \mathscr{D}[\theta^{\pm}, 
\rho^{\pm}]\boldsymbol{\varrho}\left(\rho^\pm_i,\theta^\pm_i\right)\\
 \times \exp\left[ i \sum\limits_{C = \pm} C   \int \text{d} x\: \mathscr{L}(\rho^C,\theta^C) +   J_\rho^C \rho^C +  J_\theta^C \theta^C\right].
\end {multline}
where $dx \equiv \text{d} t \text{d}^3 r$, the sum goes over the upper and lower Keldysh contours, the factor $C$ is $(\pm 1)$ for the upper/lower  contour.

The simplest observables are the mean fields, that is the expectation values of the fields. Writing
\begin {equation}
\mathscr{Z}[J^\pm_\rho,J^\pm_\theta] =  e^{i W[J^\pm_\rho,J^\pm_\theta]},
\end {equation}
the mean fields are generated by using Eq.~\eqref{pm_observable} and Eq.~\eqref{keldysh_unitary}
\begin {subequations}
\begin {align}
\rho_0^{\pm} := \langle\rho^\pm \rangle = \pm\frac{\delta W}{\delta J_\rho^\pm}\bigg\rvert_{J_\rho^\pm = 0} \mbox{ and }
\theta_0^{\pm} := \langle\theta^\pm \rangle = \pm\frac{\delta W}{\delta J_\theta^\pm}\bigg\rvert_{J_\theta^\pm = 0}.
\end {align}
\end {subequations}

Now, we can separate the system into a classical background and quantum excitations around it as follows. Suppose that the mean fields are given.  Then one can express the sources in terms of the mean fields by constructing the \textit{effective action}, that is the Legendre transform 
\begin {equation}
\label{effective_action}
\Gamma[\rho_0^\pm, \theta_0^\pm] = W[J_\rho^\pm,J_\theta^\pm] -  \sum\limits_{C = \pm} C \int \text{d} x \left( J_{\rho}^C \rho_0^C + J_\theta^C \theta_0^C \right).   
\end {equation}
Now, the sources can be expressed as
\begin {subequations}
\label{sources}
\begin {align}
\frac{\delta \Gamma}{\delta \rho_0^\pm} = \mp J_\rho^\pm \mbox{ and }
\frac{\delta \Gamma}{\delta \theta_0^\pm} = \mp J_\theta^\pm.
\end {align}
\end {subequations}
Exponentiating the effective action and
using \eqref{effective_action} one can eliminate the sources in \eqref{fullZ} in favor of the mean fields. If we define the deviations from the mean fields:
\begin {subequations}
\begin {align}
\tilde{\rho}^\pm  &= \rho^\pm - \rho_0^\pm,\\
\phi^\pm  &= \theta^\pm - \theta_0^\pm,
\end {align}
\end {subequations}
we can write the following integral equation for the effective action
\begin {multline}
\label{integral_Eq}
\!\!\!\! e^{i \Gamma[J^\pm_\rho,J^\pm_\theta]} = \int \mathscr{D}[\theta^{\pm}, 
\rho^{\pm}]\boldsymbol{\varrho}\left(\rho^\pm_i,\theta^\pm_i\right)\\
\times \exp\left[i \sum\limits_{C = \pm} C \int \text{d} x\: \mathscr{L}(\rho^C,\theta^C) - \frac{\delta \Gamma}{\delta \rho_0^C} \tilde{\rho}^C -  \frac{\delta \Gamma}{\delta \theta_0^C} \phi^C \right].
\end {multline}
The effective action, $\Gamma$, can be solved for iteratively and be expressed as a series (we restore the Planck constant below to emphasize the semiclassical nature of the expansion)
\begin {equation}
\label{loop_expansion}
\Gamma = \Gamma_0 + \hbar \Gamma_{ph} + \hbar^2 \Gamma_2 + \ldots
\end {equation}
the classical action
\begin {equation}
S[\rho^\pm, \theta^\pm] = \int  \text{d}x \: \mathscr{L}(\rho^\pm, \theta^\pm),
\end {equation}
 being the zeroth term:
\begin {equation}
\label{Gamma0}
\Gamma_0 = S[\rho_0^+ , \theta_0^+] - S[\rho_0^- , \theta_0^-].
\end {equation}
In this paper we  consider only the first order or one-loop correction $\Gamma_{ph}$ to the classical action. This correction encapsulates the quantum field of Bogoluibov quasiparticles over the background, that become phonons at long wavelengths (hence we use the subscript $ph$, a shorthand for ``\textit{phonon}'' corresponding to the first loop correction).

We substitute the lowest-order expression(~\ref{Gamma0}) into $\Gamma$'s on the right hand side of Eq.~\eqref{integral_Eq}.  On the left-hand side, we substitute $\Gamma_0 + \Gamma_{ph}$, where $\Gamma_{ph}$ is an unknown. Expanding $S$ around $\rho_0$ and $\theta_0$, and matching the terms in the equation, we find the phonon effective action $\Gamma_{ph}$  i.e. the leading-order correction in $\tilde{\rho}$ and $\phi$ to $\Gamma$.   
Defining 

$$\boldsymbol{\tilde{\varrho}} \left(\tilde{\rho}^\pm_i, \phi^\pm_i\right) = 
\boldsymbol{\varrho} \left({\rho}^\pm_i - \rho_0^\pm,\theta^\pm_i -\theta_0^\pm\right),$$
we write
\begin {multline}
\label{phonon_action}
e^{i \Gamma_{ph}[\rho_0^\pm, \theta_0^\pm]}=  \int \mathscr{D}[\tilde{\rho}^{\pm}, \phi^{\pm}]\boldsymbol{\tilde{\varrho}} \left(\tilde{\rho}^\pm_i, \phi^\pm_i\right)
\\\times 
\exp\left\{i \sum_{C=\pm}C \int dx\: \mathscr{L}^{C(2)} (\tilde{\rho}^C,\phi^C)  \right\},
\end {multline}
where the $\mathscr{L}^{2}$, depends on the path index $C = \pm$ not only through its arguments but explicitly as 
\begin {equation}
\label{L2_def}
 \mathscr{L}^{\pm(2)} (\tilde{\rho}, \theta)= (\tilde{\rho}, \phi) \cdot \frac{\delta^2 \mathscr{L}}{\delta^2 (\rho, \theta)}\bigg\vert_{\rho^\pm_0,\theta^\pm_0} \cdot (\tilde{\rho}, \phi)
\end {equation}
in multi-index notation. 
So far, the results are completely general. Now we  use the explicit Lagrangian (\ref{Lagrange}) and obtain the phonon effective Lagrangian, $\mathscr{L}^{(2)}$. After suppressing the time-path index $C=\pm$, it is 
\begin {equation}
\label{L2}
-\mathscr{L}^{(2)} = \tilde{\rho} (\partial_t + \vec{v}_0\cdot \nabla) \phi + \frac{1}{2} \tilde{\rho} \left( \frac{c_0^2}{\rho_0} + \hat{K}_Q \right) \tilde{\rho} + \frac{1}{2} \rho_0 (\nabla \phi)^2,
\end {equation}
where we defined the ``quantum pressure'' operator,
\begin {equation}
\label{quantum_pressure}
\hat{K}_Q = \frac{1}{4 m^2} \left[ \frac{(\nabla \rho_0)^2}{\rho_0^3}- \frac{\nabla \rho_0}{\rho_0^2} \cdot \nabla - \frac{\nabla^2}{\rho_0}\right].
\end {equation}


\subsection{Covariant phonon action}
The one-loop correction can be computed exactly in the strong interaction limit, as the path integral reduces to a Gaussian integral. As noted in Sec.~\ref{sec:vacuum}, at energy scales below $E_{Lorentz}$, the operator $\hat{K}_Q$ in~\eqref{L2}, that results from quantum pressure, can be neglected compared with the term $c_0^2/\rho_0$. This yields the following Lagrangian
\begin {equation}
\label{omission}
\mathscr{L}^{(2)}_{\hat{K}_Q\to 0} \to -\tilde{\rho} D_t \phi - \frac{1}{2} \tilde{\rho} \left( \frac{c_0^2}{\rho_0}  \right) \tilde{\rho} - \frac{1}{2} \rho_0 (\nabla \phi)^2, 
\end {equation} 
where we defined the material derivative
$$
D_t \phi = \partial_t \phi  + \vec{v}_0 \cdot \nabla \phi.
$$

Now, the density fluctuations, $\tilde{\rho}$, can be integrated out. To do the path integral over $\tilde{\rho}$, we can think of space-time as divided into cubes with volume $\xi^4/c_e $ and discretize the integral. Note that the integrand is a diagonal matrix  over space-time and the path integral reduces to a product of Gaussian integrals. At this point, we shorten the notations for the mean-field parameters $\rho_0$, $\theta_0$ and $\vec{v}_0$, writing them as simply $\rho$, $\theta$ and $\vec{v}$ for the sake of brevity. If we define the density matrix as
\begin {equation}
\label{phonon_density_operator}
\boldsymbol{\varrho}_{\phi}\left( \phi^+_i , \phi^-_i  \right) = \boldsymbol{\tilde{\varrho}}\left( \frac{\rho}{c^2} D_t\phi^+_i, \phi^+_i ; \frac{\rho}{c^2} D_t\phi^-_i ,\phi^-_i  \right),
\end {equation}
integrating out the $\tilde{\rho}$ field produces:
\begin {equation}
\label{phonon_partition}
Z_{ph} = e^{i \Gamma_{ph}[\rho^\pm, \theta^\pm]}=  \int \mathscr{D}'[\phi^{\pm}]\boldsymbol{\varrho}_{\phi}
e^{i\left(S^+_{ph}[\phi^+] -  S^-_{ph}[\phi^-] \right)}.
\end {equation}
with the measure of the path integral being, again suppressing the $\pm$ signs,
\begin {subequations}
\label{measure}
\begin {align}
&\mathscr{D}' [\phi] = \prod d\phi \sqrt{\frac{\rho}{2 \pi c^2}},
\end {align}
\end {subequations}
and the following covariant action for phonons
\begin {equation}
\label{matter}
S_{ph} = \frac{1}{2} \int \text{d} t\: \text{d}^3 x \: \frac{\rho}{c^2} \left[ \left(D_t \phi \right)^2 - c^2\left(\nabla \phi\right)^2  \right].
\end {equation}

This action can also be obtained through a classical treatment of the Lagrangian Eq.~\eqref{Lagrange}.~\cite{Stone} The material derivative $D_t\phi = \partial_t \phi + \vec{v}\cdot \nabla \phi$  is the measure of the time rate of change of $\phi$ in a frame comoving with the fluid. This means the action in Eq.~\eqref{matter} describes non-dispersive waves with speed $c$ in the fluid comoving frame. Galilean invariance of the fluid system requires that the sound wave velocity $\vec{u} = d \vec{x}/dt$ in the lab frame satisfy $(\vec{u} - \vec{v}) = c^2$. This means the sound rays with velocity with $d \vec{x}/dt = \vec{u}$ are null rays  on the manifold with line element

\begin {equation}
\label{line_element}
ds^2 = -\frac{\rho}{ c} [-(c^2 - v^2) dt^2 -2 \vec{v}\cdot \vec{dx} dt + \vec{dx}\cdot \vec{dx}].
\end {equation} 

This is the line element for any analogue gravity system with background Galilean symmetry up to a conformal factor, for which the choice of $\rho/c$ allows us to write the phonon action of Eq~\eqref{matter} in the following suggestive form~\cite{Unruh,Stone}

\begin {equation}
\label{phonon_action}
S_{ph} =\frac{1}{2} \int \text{d}^4 x \sqrt{-g} g^{\mu\nu} \partial_\mu \phi \partial_\nu \phi.
\end {equation}
Here, the metric can be read off from the line element Eq.~\eqref{line_element} by writing $ds^2 = g_{\mu\nu} dx^{\mu} dx^{\nu} $ as
\begin {equation}
\label{metric}
g_{\mu\nu} = \frac{\rho}{c}
\begin{pmatrix}
c^2 - v^2 & \vec{v}^T\\
\vec{v} & -\mathbf{I}_{3\times 3}
\end{pmatrix},
\end {equation}
here $(+ - - -)$ convention is used. 

The volume measure factor turns out to be $\sqrt{-g} = \rho^2/c $ . At static equilibrium the line element Eq.~\eqref{line_element} is that of Minkowski space.

The measure in Eq.~\eqref{measure} is written as
\begin {equation}
\label{anom_measure}
\mathscr{D}' [\phi] = \prod_x \underbrace{(g^{00} )^{1/2}}_{\small \mbox{``anomaly''}} (-g)^{1/4} d\phi.
\end {equation}
We note that this measure is manifestly non-covariant due to the coordinate dependent factor  $g^{00}$. For a field in curved space-time the covariant measure ought to be $(-g)^{1/4}$.~\cite{Toms_measure, hawking1977}. This means, although the phonon action Eq.~\eqref{matter} is covariant, the path integral is not, leading to a quantum anomaly. We will come back to this issue in Sec.~\ref{sec:covariant_conservation} where we derive the conservation law of the stress tensor.

\begin{figure}[h]
\begin{center}$
\begin{array}{cc}
\includegraphics[width=0.45\textwidth]{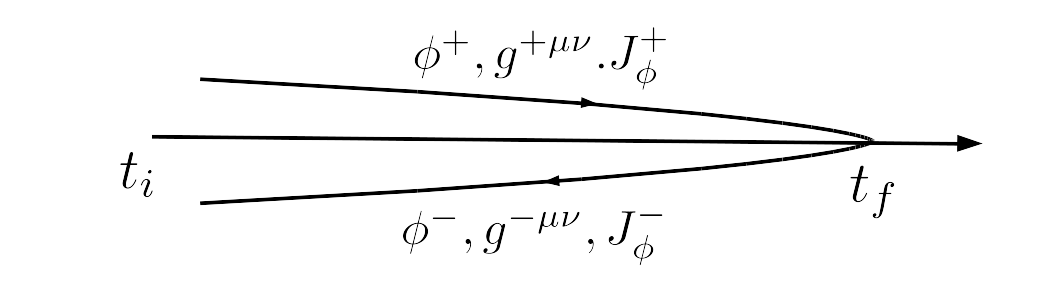}
\end{array}$
\end{center}
\caption{The Keldysh contour for the phonon field. The metric tensor depends on the background variables $\rho_0^\pm$ and $\theta_0^\pm$ and therefore is also defined twice.}
\label{Fig2}
\end{figure}

Before going into obtaining equations of motion from the effective action of the Keldysh field theory, it is worthwhile to list a number of properties that Keldysh theory obeys.
We draw Fig.~\ref{Fig2} to show the forward/backward fields and sources of the phonon field.  We identified Eq.~\eqref{phonon_partition} with the closed-time-path partition function of phonons $Z_{ph}$. The first order correction $\Gamma_{ph}$ to the classical action is then given as:
\begin {equation}
\Gamma_{ph} = -i \log Z_{ph}[g^+ , g^-].
\end {equation}
As a consequence of unitarity, similar to Eq.\eqref{keldysh_unitary},
\begin {equation}
\label{unitarity}
Z_{ph}[g^\pm = g] = 1,
\end {equation}
i.e. when the background is the same on the forward and backward directions, the product of backward and forward evolution operators is unitary.  Also, from Eq.~\eqref{phonon_partition},
\begin {equation}
\label{complex}
Z_{ph}[g^+,g^-; J^+, J^-] = (Z_{ph}[g^-,g^+; J_{\phi}^-, J_{\phi}^+])^*,
\end {equation}
where $J_{\phi}^\pm$ are additional sources attached in order to compute expectation values of $\phi$. 
It follows from Eq.~\eqref{unitarity} and Eq.~\eqref{complex} that the effective action satisfies
\begin {subequations}
\label{gamma_unitary}
\begin {align}
\Gamma_{ph} [g, g] &= 0,\\
\Gamma_{ph} [g^+, g^-] & = -\Gamma_{ph}^* [g^-, g^+] .
\end {align}
\end {subequations}
These properties are handy while computing the Keldysh correlation functions. 

\section{analogue Einstein equations and two-fluid hydrodynamics}\label{sec:two-fluid}

In general relativity, the relationship between matter and space-time curvature is governed by the Einstein's equation. Being covariant under coordinate transformations, the matter field obeys the covariant conservation law. However, this is not a total conservation law, as there is energy momentum exchange between fields and the space-time curvature. There are pseudo-tensor constructions like Einstein pseudotensor or the Landau-Lifshitz pseudotensor that quantify the stress energy of the gravitational field.~\cite{Horski_conservation, Landau_gravity} These pseudo-tensors, when added to the stress tensor of matter, become a totally conserved quantity.

In this section, in analogy with general relativity, we will start with the first loop effective action
\begin {equation}
\label{first_loop}
\Gamma = S[\rho^+ , \theta^+] - S[\rho^- , \theta^-] - i \log Z_{ph}[g^\pm],
\end {equation}
where the metric $g^{\pm}$ is a functional of $\rho^\pm$ and $\theta^\pm$ according  to Eq.~\eqref{metric}. The effective action Eq.~\eqref{first_loop} is in agreement with the one postulated in Ref.~[\citen{Balbinot}] to compute backreaction corrections to acoustic black holes where the quantum effects, i.e. Hawking radiation, is important.
We will first write the stress-energy tensor for the covariant phonons starting from the effective action Eq.~\eqref{effective_action}. Then we will write down an analogue Einstein equation that describes the evolution of metric tensor~\eqref{metric} due to the stress-energy of phonons, which play the part of matter. To complete the analogy, we will derive a total conservation law by using the covariant conservation law and the analogue Einstein's equation. Moreover, we will show that the conserved quantity is a canonical Noether current and therefore describes the total conservation of momentum and energy in the lab frame. This means two fluid hydrodynamics \textit{directly} follows from the analogue gravity formalism.

\subsection{Hilbert Stress-Energy operator of the covariant phonon field}
\label{sec:stress}
The expectation value of the stress-energy operator can be defined by using Schwinger's variational principle~\cite{birrell1984quantum} as
\begin {multline}
\label{stress}
\left\langle \hat{T}_{\mu\nu}\right\rangle = -i \frac{2}{\sqrt{-g}} \frac{\delta \log Z_{ph}}{\delta (g^+)^{\mu\nu}} \bigg\rvert_{g^+ = g^-} =  \\ \frac{2}{\sqrt{-g}} \frac{\delta \Gamma_{ph}}{\delta (g^+)^{\mu\nu}} \bigg\rvert_{g^+ = g^-}.
\end {multline}
Then, the stress-energy operator is  defined (symmetrized for convenience) as 
\begin {multline}
\label{stress_tensor}
\hat{T}_{\mu\nu}(x) = \frac{1}{2}\lbrace\partial_\mu \hat{\phi}(x), \partial_\nu \hat{\phi}(x)\rbrace -\frac{1}{2} g_{\mu\nu} \partial^\alpha \hat{\phi}(x) \partial_\alpha \hat{\phi}(x).
\end {multline}
This expression is problematic because it contains a product of field operators at the same space-time point and generally leads to divergences. These can be cured through a variety of regularization and renormalization schemes.~\cite{birrell1984quantum}  One of these methods is called point splitting where the field operators are taken on different space time points, and the limit of coincidence is taken after performing derivatives and averages. In semiclassical gravity diverging quantities are renormalized into coupling constants in Einstein's equation.~\cite{birrell1984quantum,Hu_Verdaguer_primer} 

Physically, the zero point quantum fluctuations that add up to an infinite vacuum energy. In the strongly interacting analoguesystem, the divergent quantities are believed to be already accounted for in the background as a part of the internal energy of the fluid.~\cite{Volovik_book}  This ensures the stability of the liquid droplets, by renormalizing the equilibrium pressure to zero. Recently, the role of zero point energy in the formation of stable macroscopic droplets in strongly interacting BEC's was investigated both theoretically~\cite{Petrov} and experimentally.~\cite{macrodroplet}  Therefore, assuming that the vacuum energy is already renormalized into the background energy, we will formally discard the divergent piece of the stress-energy expectation value. Note that, in field theory on flat space-time, the divergence is tacitly discarded through the normal ordering of operators. 

Let $ \langle \hat{T}_{\mu\nu}(x)\rangle_{div}$ represent the divergent piece in the expectation value for the stress-energy operator. Throughout the paper we will refer to the renormalized stress-energy as:
\begin {multline}
\label{stress_renorm}
\left\langle \hat{T}_{\mu\nu}(x)\right\rangle = \frac{1}{2}\left( \delta^\alpha_\mu \delta^\beta_\nu + \delta^\alpha_\nu \delta^\beta_\mu- g_{\mu\nu}g^{\alpha\beta}\right) \times \\
\lim_{x',x'' \to x} \frac{\partial}{\partial x'^\alpha} \frac{\partial}{\partial x''^\beta} \left\langle \mathbb{T} \hat{\phi}(x') \hat{\phi}(x'') \right\rangle\  - \left\langle \hat{T}_{\mu\nu}(x)\right\rangle_{div}.
\end {multline}
where $\mathbb{T}$ is the time ordering operator. The time ordered correlation function  is equivalent to the forward-forward correlation function of the Keldysh theory~\cite{Kamenev_book}
\begin {multline}
\left\langle \mathbb{T} \hat{\phi}(x') \hat{\phi}(x'') \right\rangle\ 
=\left\langle \hat{\phi}^+(x') \hat{\phi}^+(x'') \right\rangle 
\\= \frac{\delta^2 Z_{ph}[J^+, J^-]}{\delta J^+(x') \delta J^+(x'') } \bigg|_{J^+ = J^- = 0}. 
\end {multline}

\subsection{Semiclassical analogue Einstein equations and the ``phonon matter''}

Having defined the stress tensor of the matter field (phonons), we now write down the equations of motion for the superfluid and the phonons and argue that it is analogous to the semi-classical Einstein's equation.

After dropping the quantum pressure, and using Eq.~\eqref{stress}, the Euler-Lagrange equations that follow from the effective action Eq.~\eqref{first_loop} in the limit $(\rho^\pm$, $\theta^\pm) \to (\rho$, $\theta)$  are the fluid equations of motion Eq.~\eqref{GPE_hydro} with semiclassical source terms
\begin {subequations}
\label{euler}
\begin {align}
\label{cont}
\partial_t \rho + \nabla \cdot(\rho \vec{v}) &= \frac{1}{2} \nabla \cdot \left(\sqrt{-g} \left\langle \hat{T}_{\mu\nu} \right\rangle  \frac{\delta g^{\mu\nu}}{\delta \vec{v}}\right),\\
\label{Bernouilli}
\partial_t \theta + \frac{1}{2} v^2 + \mu(\rho)  &= \frac{1}{2} \left(\sqrt{-g} \left\langle \hat{T}_{\mu\nu} \right\rangle  \frac{\delta g^{\mu\nu}}{\delta \rho}\right).
\end {align}
\end {subequations}
Here $\mu(\rho) = \partial \varepsilon/\partial \rho$ is the local chemical potential for the superfluid. 

Given the initial density operator of phonons and initial 
values and boundary conditions for the background fields $\rho$ and $\theta$, one can compute the metric everywhere by solving Eq.~\eqref{euler} and plugging the solutions into the definition Eq.~\eqref{metric}. The source terms on the right hand side must be self-consistent with this solution, as the metric appears both explicitly in \eqref{euler} and also inside the definition of the stress tensor in Eq.~\eqref{stress_renorm}. This is because the metric tensor determines how the sound field propagates, classically expressed as Eq.~\eqref{cov_wave}. In this respect, the Eqs.~\eqref{euler} resembles Einstein's equations where the Eulerian left hand sides provide dynamics to the metric tensor and are analogous to the Einstein tensor and the stress-tensor corresponds to the matter whose motion is dictated by the curvature.

\subsection{Canonical versus Covariant Conserved Currents }
\label{sec:covariant_conservation}
Here, we show that the covariant conservation law and the analogue Einstein equations Eq.~\eqref{euler} leads to a canonical conservation law for energy-momentum.

Classically, the stress tensor in Eq.~\eqref{stress}, obeys the covariant conservation law
\begin {multline}
\label{cov_con}
\nabla_\mu \tensor{T}{^\mu_\nu}  =  \partial_\mu \tensor{T}{^\mu_\nu} +  \tensor{{T}}{^\theta_\nu}  \tensor{\Gamma}{^\mu_\theta_\mu} -   \tensor{T}{^\mu_\theta}  \tensor{\Gamma}{^\theta_\mu_\nu}
 = 0,
\end {multline}
owing to the 
fact that it is derived from the effective action,\cite{birrell1984quantum}
where $\nabla_{\mu}$ denotes the covariant derivative. Writing the definitions of Christoffel symbols Eq.~\eqref{chris_def} in terms of the metric and using Eq.~\eqref{chris_1} and Eq.~\eqref{chris_2}, we get:
\begin {multline}
\label{cov_con2}
\nabla_\mu \tensor{T}{^\mu_\nu}  = 
\frac{1}{\sqrt{-g}}  \partial_{\mu}\left( \tensor{T}{^\mu_\nu}  \sqrt{-g}\right) +\frac{1}{2}\tensor{T}{_\alpha_\beta} \partial_{\nu}g^{\alpha \beta} = 0.
\end {multline}
The partial derivative is streamlined with comma notation whenever convenient, i.e. for any quantity $A$, $\partial _\mu A = A_{,\mu}$.
The Lagrangian density for phonons can be extracted from Eq.~\eqref{phonon_action} as
\begin {equation}
\mathscr{L}_{ph} = \frac{1}{2}\sqrt{-g} g^{\mu\nu} \partial_\mu \phi \partial_\nu \phi.
\end {equation}
Then the first term in Eq.~\eqref{cov_con2} is related to the canonical stress tensor of phonons, and by inspecting the classical version of the Hilbert stress-energy tensor Eq.~\eqref{stress_tensor}, it can be written as
\begin {equation}
\label{canon_stress}
 \sqrt{-g} \tensor{T}{^\mu_\nu} =  \frac{\partial \mathscr{L}_{ph} }{\partial \phi_{,\mu}} \:\phi_{,\nu}
-\mathscr{L}_{ph} \delta{^\mu}{_\nu}.
\end {equation}
By using the chain rule and the definition Eq.~\eqref{stress}, the second piece in Eq.~\eqref{cov_con2} reduces to
\begin {equation}
\label{second_piece}
\frac{1}{2}\sqrt{-g}\tensor{{T}}{_\alpha_\beta} \partial_{\nu}g^{\alpha \beta}  = \frac{\partial \mathscr{L}_{ph}}{\partial \theta_{,\mu}}\theta_{,\mu\nu} + \frac{\partial \mathscr{L}_{ph}}{\partial \rho} \rho_{,\nu}.
\end {equation}

Let $\mathscr{L}_{cl}$ denote the Lagrangian density of the background after the quantum pressure term is dropped
\begin {equation}
-\mathscr{L}_{cl} = \rho\partial_t \theta + \frac{1}{2}\rho \vec{v}^2 + \varepsilon(\rho).
\end {equation}
By using the equations of motion Eq.~\eqref{euler}, in the Euler-Lagrange form, as shown in Appendix~\ref{app:conservation} write can write Eq.~\eqref{cov_con2} as the conservation of the following current
\begin {multline}
\label{Noether}
\tensor{\mathfrak{T}}{^\mu_\nu} = \frac{\partial\mathscr{L}_{ph}  }{\partial \phi_{,\mu}} \phi_{,\nu}
+ \frac{\partial (\mathscr{L}_{ph}+\mathscr{L}_{cl})}{\partial \theta_{,\mu}} \:  \theta _{,\nu} -\left( \mathscr{L}_{ph}+\mathscr{L}_{cl}\right) \delta{^\mu}{_\nu}.
\end {multline}

If we define the total Lagrangian of the background-phonon composite system
$$
\mathscr{L}_{sys} = \mathscr{L}_{cl} + \mathscr{L}_{ph},
$$
and noticing that $\partial \mathscr{L}_{cl}/ \partial \rho_{,\mu}$ and $\partial \mathscr{L}_{cl}/ \partial \phi_{,\mu}$ both vanish, $\mathfrak{T}$ in Eq.~\eqref{Noether} is 
 is precisely the conserved 
Noether current
$$
\tensor{\mathfrak{T}}{^\mu_\nu} = \frac{\partial\mathscr{L}_{sys}  }{\partial (\rho,\theta,\phi)_{,\mu}} \cdot  (\rho,\theta,\phi)_{,\nu} - \mathscr{L}_{sys} \delta{^\mu}{_\nu}.
$$ 
due to the space time translation invariance of the overall system represented by  the 
effective action Eq.~\eqref{first_loop}. 

We define $\mathcal{T}$, the stress energy tensor of the analoguegravitational field as
\begin {equation}
\label{grav}
\sqrt{-g}\tensor{\mathcal{T}}{^\mu_\nu} = \frac{\partial (\mathscr{L}_{ph}+\mathscr{L}_{cl})}{\partial \theta_{,\mu}} \:  \theta _{,\nu} -\mathscr{L}_{cl} \delta{^\mu}{_\nu}.
\end {equation}
This expression generates the familiar energ and momentum density in the background. For example the energy density of the background in the laboratory frame follows from Eq.~\eqref{grav} as 
\begin {equation}
\sqrt{-g}\tensor{\mathcal{T}}{^0_0} = \frac{1}{2}\rho \vec{v}^2 + \varepsilon(\rho).
\end {equation}
Similarly, the laboratory frame momentum density of the background follows from Eq.~\eqref{grav} as
\begin {equation}
\sqrt{-g}\tensor{\mathcal{T}}{^0_i} = -\rho v_i.
\end {equation}

The Noether current Eq.~\eqref{Noether} can be written as the current due to the background and the excitations as 
Eq.~\eqref{cov_con} as the
\begin {equation}
\label{einstein_con}
\tensor{\mathfrak{T}}{^\mu_\nu} = \sqrt{-g}\tensor{T}{^\mu_\nu} + \sqrt{-g}\tensor{\mathcal{T}}{^\mu_\nu}.
\end {equation}
This means the mixed canonical stress tensor of the 
excitations correspond to energy-momentum corrections due to  excitations in the laboratory frame.

However, because of the 
covariance anomaly of the quantum phonon field that manifests itself in the path integral measure 
Eq.~\eqref{anom_measure}, the covaraint derivative of the expectation value of the quantum stress 
operator must be equal to an anomalous current, i.e. $\nabla_{\mu}\langle \tensor{\hat{T}}{^\mu_\nu} \rangle = J_{\nu}^{anom}$. Furthermore this current should be Galilean covariant due to the overall Galilean invariance of the system.  This is the second type of anomaly in the analogue gravity system, the first being the trace anomaly due to Hawking radiation, which occurs when there is a sonic horizon in the system.~\cite{Hawking_BEC}  In this paper we assume that no sonic horizon exists and that the quantum pressure term is weak everywhere. We defer the rigorous analysis and computation of the
anomalous current to another publication. Since anomalies are necessarily quantum effects, in the regimes we work, they should be washed out by thermal contributions. Then for the expectation
\begin {equation}
\label{einstein_quant_con}
\tensor{\mathfrak{T}}{^\mu_\nu} = \sqrt{-g} \left\langle \tensor{\hat{T}}{^\mu_\nu}\right\rangle + \sqrt{-g}\left\langle \tensor{\mathcal{T}}{^\mu_\nu}[ \hat{\phi}]\right\rangle,
\end {equation}
the  conservation law 
\begin {equation}
\label{Noether_con}
\partial_\mu \tensor{\mathfrak{T}}{^\mu_\nu} =  0
\end {equation}
holds.
In the next section we rewrite Eq.~\eqref{Noether_con} and Eq.~\eqref{euler} in the two-fluid variables. 

\subsection{Mass-energy-momentum balance and the covariant conservation of stress-energy operator}

In this section, we complete the connection of the analogue Einstein equation Eq.~\eqref{euler} and the covariant conservation law Eq.~\eqref{cov_con2} and two-fluid hydrodynamics by identifying the normal and superfluid components in terms of the background fields and the stress tensor. 

Inspecting the conservation law Eq.~\eqref{einstein_quant_con} that we derived in Sec.~\ref{sec:covariant_conservation},  we define the momentum density $\vec{P}$,  energy density $E$ and energy flux $\vec{Q}$  due to phonons, in the lab frame
\begin {subequations}
\label{dictionary}
\begin {align}
P_i:&=-\sqrt{-g} \left\langle \tensor{\hat{T}}{^0_i}\right\rangle,  \\
 E :&= \sqrt{-g}  \left\langle \tensor{\hat{T}}{^0_0}  \right\rangle, \\
 Q_i :&= \sqrt{-g}  \left\langle \tensor{\hat{T}}{^i_0}  \right\rangle.
\end {align}
\end {subequations}
Here, the quantities on the left hand side are momentum density, momentum flux tensor and energy density.  Being Cartesian tensors, their indices are uppered/lowered by using the Kronecker delta function. By Galilean transforming the quantities on the right hand side according to the usual tensor transformation rules, we observe that $P_i$ is Galilean invariant while $E$ and $Q_i$ are not.

We define the following symmetric, Galilean invariant, Cartesian momentum flux tensor
\begin {equation}
\pi_{i j}:= \frac{c}{\rho}\sqrt{-g} \left\langle \tensor{\hat{T}}{_i_j}  \right\rangle = \frac{c}{\rho}\sqrt{-g} \left\langle \tensor{\hat{T}}{^\mu_j}  \right\rangle g_{i\mu}.
\end {equation}
By using the expression for the metric Eq.~\eqref{metric} and the definitions in Eq.~\eqref{dictionary}, the mixed momentum flux tensor due to phonons is
\begin {equation}
\label{mixed_mom}
- \sqrt{-g} \left\langle \tensor{\hat{T}}{^i_j}  \right\rangle =\pi_{i j} + P_j v_i^s , \\
\end {equation}

The equations of two fluid hydrodynamics are conservation laws for the density, momentum and energy for the superfluid  and normal components of the system denoted by the superscripts $s$ and $n$ respectively. We  make the following identification of velocity and density in Eq.~\eqref{euler} as the velocity of the superfluid component and the total density (i.e. superfluid plus normal)respectively.
\begin {subequations}
\begin {align}
\vec{v} &:= \vec{v}_s,\\
\rho &:= \rho_s + \rho_n.
\end {align}
\end {subequations}
Consequently, $\vec{v}_s = \nabla \theta$. 

Now, we identify the analogue Einstein equations Eq.~\eqref{euler} as mass conservation and superflow equations respectively, by writing the phonon contribution in terms of energy and momenta defined in Eq.~\eqref{dictionary}.
 
The inverse metric in Eq.~\eqref{euler} follows from Eq.~\eqref{metric} as
\begin {equation}
\label{incv_metric}
g^{\mu\nu} = \frac{1}{\rho c}
\begin{pmatrix}
1 & \vec{v}_s^T\\
\vec{v}_s & -c^2\mathbf{I}_{3\times 3} + \vec{v}_s \vec{v}_s^T
\end{pmatrix}.
\end {equation}

If $A$ is any tensor with two covariant indices contracted and $\kappa = \partial \ln c/ \partial \ln\rho$ characterizes the  logarithmic derivative of the energy of a linearly dispersing sound wave with respect to the density of the medium, then the derivative of the metric Eq.~\eqref{incv_metric} obeys the following rules
\begin {subequations}
\begin {align}
\label{contract}
\frac{\delta g^{\mu\nu}}{\delta \rho}\tensor{A}{_{\mu\nu}_{...}^{...}} &= \frac{1}{\rho}(\kappa -1) \tensor{A}{^\mu_\mu_{...}^{...}} - 2 \kappa c \tensor{A}{^0^0_{...}^{...}},\\
\frac{\delta g^{\mu\nu}}{\delta v_s^i}\tensor{A}{_{\mu\nu}_{...}^{...}} &= \tensor{A}{^0_i_{...}^{...}} +  \tensor{A}{_i^0_{...}^{...}}.
\end {align}
\end {subequations}

The source current can be written as quasi-particle momentum by making the identification
\begin {multline}
\label{momentum_density}
-\frac{1}{2} \sqrt{-g} \left\langle \hat{T}_{\mu\nu} \right\rangle  \frac{\delta g^{\mu\nu}}{\delta \vec{v}_s} = -\sqrt{-g} \left\langle \tensor{\hat{T}}{^0_i}\right\rangle  = P_i.
\end {multline}
Then the first analogue Einstein equation Eq.~\eqref{cont} is recast as the continuity equation for mass
\begin {equation}
\label{mass_cont}
\partial_t \rho + \nabla \cdot (\rho \vec{v}_s + \vec{P}) = 0.
\end {equation}

The source term in Eq.~\eqref{Bernouilli} reads
\begin {multline}
\label{energy_density}
\frac{1}{2} \sqrt{-g} \left\langle \hat{T}_{\mu\nu} \right\rangle  \frac{\delta g^{\mu\nu}}{\delta \rho}\\= \frac{\sqrt{-g}}{2 \rho}[\kappa -1] \left\langle \tensor{\hat{T}}{^\mu_\mu} \right\rangle_{anom}  - \frac{\kappa}{\rho} \rho c \sqrt{-g} \left\langle \hat{T}^{0 0} \right\rangle    \\:=-\frac{\kappa}{\rho} [E - \vec{v}_s \cdot \vec{P}] =  -\frac{\kappa}{\rho} E_{0} = -\frac{\partial E_{0}}{\partial \rho}.
\end {multline}
Here, the trace of the stress-operator is zero in the absence of an acoustic horizon where the Hawking radiation creates a trace anomaly. The frame shifted $E_{0}$, by virtue of its tensorial expression being manifestly Galilean invariant, is the comoving frame energy density of quasiparticles. In these variables the Bernoulli  Eq.~\eqref{Bernouilli} takes its familiar form
\begin {equation}
\label{london}
\partial_t \theta + \frac{1}{2} v_s^2 + \mu (\rho) + \frac{\partial E_0}{\partial \rho} = 0. 
\end {equation}
A corrolary of this equation is that the classical Lagrangian $\mathscr{L_{cl}}$ of the background is related to the isotropic pressure in the system
\begin {multline}
\label{pressure}
\mathscr{L}_{cl}  = -\rho\partial_t \theta - \rho\frac{1}{2} v_s^2 - \varepsilon(\rho) \\= [\rho \mu(\rho) - \varepsilon(\rho)] + \rho\frac{\partial E_0}{\partial \rho} - E_0 + E_0  = p + E_0 .
\end {multline}

From Eq.~\eqref{Noether_con}, the momentum conservation equation reads
\begin {equation}
\label{momentum_balance}
\partial_\mu \tensor{\mathfrak{T}}{^\mu_i} = \partial_t \tensor{\mathfrak{T}}{^0_i} + \partial_j \tensor{\mathfrak{T}}{^j_i} =0.
\end {equation}
Writing $\mathfrak{T}$ as in Eq.~\eqref{einstein_quant_con}, computing $\mathcal{T}$ according to Eq.~\eqref{grav} and using Eqs.~\eqref{dictionary}, Eq.~\eqref{mixed_mom} and Eq.~\eqref{pressure} we get
\begin {multline}
\label{mom_cont}
\!\!\!\!\!\partial_t (\rho v_{si} + P_i) + \partial_j([\rho v_{si} + P_i] v_{sj} + P_j v_{si} + \pi_{i j} ) + \partial_i (p +E_0) = 0.
\end {multline}
where repeated indices are summed.

Similarly, From Eq.~\eqref{Noether_con}, the energy conservation equation reads
\begin {equation}
\label{energy_balance}
\partial_\mu \tensor{\mathfrak{T}}{^\mu_0} = \partial_t \tensor{\mathfrak{T}}{^0_0} + \partial_j \tensor{\mathfrak{T}}{^j_0} =0.
\end {equation}
Writing $\mathfrak{T}$ as in Eq.~\eqref{einstein_quant_con}, computing $\mathcal{T}$ according to Eq.~\eqref{grav} and using Eqs.~\eqref{dictionary} and Eq.~\eqref{london}, we get
\begin {multline}
\label{energy_cont}
\partial_t\left(\frac{1}{2}\rho v^2 + \varepsilon(\rho) + E\right) \\+ \partial_i\left(Q_i + \left[\rho v_{si} + P_i\right] \left[\frac{1}{2}v_s^2  + \mu(\rho) + \partial E_0/\partial \rho\right]\right)=0
\end {multline}

In summary, the analogue Einstein equation Eq.~\eqref{euler}, the covariant conservation law of stress tensor Eq.~\eqref{cov_con} and the consequent conservation law Eq.~\eqref{Noether_con} leads to the continuity of mass Eq.~\eqref{mass_cont}, Bernouilli equation Eq.~\eqref{london} and the conservation of momentum Eq.~\eqref{mom_cont}. If we define the current density $\vec{J}$, energy density $\mathscr{E}$, momentum flux $\Pi$ and the superfluid potential $G$ as
\begin {subequations}
\begin {align}
&{J}_i := \rho v_{si} + P_i,\\
& \mathscr{E} :=\frac{1}{2}\rho v^2 + \varepsilon(\rho) + E,\\
 & \!\!\!\! \tensor{\Pi}{_i_j } := \rho v_{s i} v_{sj} + P_i v_{s j} + P_j v_{s i}  + \tensor{\pi}{_i_j} +[ p+ E_0] \tensor{\delta}{_i_j},\\
&G :=  \frac{1}{2} v_{s}^ 2 + \mu(\rho) + \frac{\partial E_{0}}{\partial \rho}.
\end {align}
\end {subequations} 
we rewrite Eq.~\eqref{mass_cont}, Eq.~\eqref{mom_cont}, Eq.~\eqref{energy_cont} and Eq.~\eqref{london} as the Landau- Khalatnikov equations for the superfluid

\begin {subequations}
\label{two-fluid}
\begin {align}
\label{mass_con}
\partial_t \rho + \nabla \cdot \vec{J} &= 0, \\
\label{mom_con}
\partial_t J_i + \nabla_j  \boldsymbol{\Pi}_{i j} &= 0,\\
\label{energy_con}
\partial_t \mathscr{E} + \nabla \cdot (\vec{Q} + \vec{J} G) &= 0,\\
\label{superfluid}
\partial_t \vec{v}_s + \nabla G &=0.
\end {align}
\end {subequations}

The Eqs.~\eqref{mass_con} and ~\eqref{superfluid} share  the manifest Galilean invariance of Euler equations, because the $P_i$ and $E_0$ are comoving frame momentum and energy of phonons. The Galilean invariance of  Eqs.~\eqref{mom_con} and ~\eqref{energy_con} follow immediately, because they are derived using the covariant conservation law Eq.~\eqref{cov_con}, that is valid in all frames, and the analogue Einstein equations, which in the two-fluid language become Eqs.~\eqref{mass_con} and ~\eqref{superfluid}.

Finally, note that the dissipative effects due to the normal fluid are taken into account in the heat flux $\vec{Q}$ and the momentum flux tensor $\pi_{i j}$. The dissipative components of these tensors can be computed according to linear response theory which we explain in the next section.  In the limit $\vec{v}_s \to 0$, the Eqs.~\eqref{two-fluid} become that of a normal fluid with $\rho \to \rho_n$ is the normal density and $P_i \to \rho_n  v_{ni}$ defines the velocity of normal component. Assuming Stoke's constitutive law for stress, the momentum balance equation~(\ref{mom_con}) becomes the Navier-Stokes equation.

\section{Stochastic analogue Einstein equations}
\label{sec:stochastic}

Starting from the Lagrangian Eq.~\eqref{Lagrange}, we first derived the Euler  equations Eq.~\eqref{GPE_hydro} for the background and improved it to contain the phonons and called it the analogue Einstein (order one in metric perturbation) equations Eq.~\eqref{euler}.
 
In this section we will go to second order in the metric perturbations and obtain dissipation and noise kernels due to phonons through a generalized linear response method.  We will write an effective stochastic action for phonons that capture this noise and derive the stochastic analogue Einstein equation. This equation when linearized around a deterministic solution for the background fields $\rho$ and $\theta$, describes the motion of the stochastic corrections to the solution. Finally we show that in thermodynamic equilibrium the stochastic forcing term is balanced by the dissipation, that is the fluctuation-dissipation relation holds.

\subsection{Linear response and the covariant stress-energy correlator}
To see the semi-classical expansion due to metric perturbations, we first define the vector in time-path space composed of forward/backward metric tensors
\begin {equation}
\vec{\mathbf{g}} = \left( g^{+\mu\nu}(x), g^{-\mu\nu}(x) \right).
\end {equation}
Then the expansion of phonon effective action up to second order in metric perturbations is
\begin {equation}
\Gamma_{ph} = \frac{\boldsymbol{\delta} \Gamma_{ph}}{\boldsymbol{\delta}\vec{\mathbf{g}}} \cdot \boldsymbol{\delta} \vec{\mathbf{g}} + \frac{1}{2} \boldsymbol{\delta} \vec{\mathbf{g}}\cdot \frac{\boldsymbol{\delta}^2 \Gamma_{ph}}{\boldsymbol{\delta} \vec{\mathbf{g}} \boldsymbol{\delta} \vec{\mathbf{g}}}\cdot {\boldsymbol{\delta}\vec{\mathbf{g}}} + O(\boldsymbol{\delta}\vec{\mathbf{g}}^3).
\end {equation}
Here the boldface symbol $\boldsymbol{\delta}$ is for variation with respect to the element $g^{+\mu\nu}$ is defined as 
\begin {equation}
\frac{\boldsymbol{\delta} \Gamma}{\boldsymbol{\delta} g^{+\mu\nu}} = \frac{1}{\sqrt{-g(x)}}\frac{\delta \Gamma[g^{+}(x),g^{-}(x)]}{\delta g^{+\mu\nu}(x)}\bigg\rvert_{g^+ = g^- = g} ,
\end {equation}
and the product of two time-path vectors $(\vec{\mathbf{A}}\cdot\vec{\mathbf{B}})$  sums over all path and tensor indices and integrates over position by using the metric $g^+ = g^- = g$, namely,
\begin {multline}
\vec{\mathbf{A}} \cdot \vec{\mathbf{B}} =\int \text{d}^4 x \sqrt{-g(x)}\bigg(A^{+\mu\nu}(x) B^{+}_{\mu\nu}(x) \\+ A^{-\mu\nu}(x)B^{-}_{\mu\nu}(x) \bigg).
\end {multline}

Using Eq.~\eqref{unitarity}, \eqref{complex} and \eqref{gamma_unitary}, it follows that the first order variation of the phonon effective action is

\begin {equation}
\label{first_order}
 \frac{\delta \Gamma_{ph}}{\delta \vec{\mathbf{g}}} = \frac{1}{2}
\begin{pmatrix}
\mathbf{T} & ,- \mathbf{T}
\end{pmatrix},
\end {equation}
where 
\begin {equation}
\mathbf{T} = \left\langle \hat{T}_{\mu\nu}(x) \right\rangle.
\end {equation} 

We define the deviations of the stress energy operator from its expectation value
\begin {equation}
 \hat{t}_{\mu\nu}(x)  = \hat{T}_{\mu\nu}(x) -  \left\langle\hat{T}_{\mu\nu}(x) \right\rangle.
\end {equation}
Following Martin and Verdaguer ~\cite{Martin_Verdaguer_semiclassical_gravity} we define the local stress deviation $\mathbf{K}$, the causal response kernel  $\mathbf{H}$  and the noise kernel $\mathbf{N}$ that are bi-tensors made of correlators of $\hat{t}$ 
\begin {subequations}
\label{kernels}
\begin {align}
\label{response}
\hat{H}_{\mu\nu\alpha\beta} (x,y) &= -i\theta(x^0 - y^0)  \left\langle  \left[ \hat{t}_{\mu\nu}(x), \hat{t}_{\alpha\beta}(y) \right]\right\rangle, \\
\label{noise}
\hat{N}_{\mu\nu\alpha\beta} (x,y) & = \frac{1}{2} \left\langle \left\lbrace \hat{t}_{\mu\nu}(x),\hat{t}_{\alpha\beta}(y) \right\rbrace\right\rangle,\\
\label{K_response}
\hat{K}_{\mu\nu\alpha\beta} (x,y) &= \frac{-4}{\sqrt{g(x)g(y)}}
\left\langle \frac{\delta^2 S_{ph}[\hat{\phi}]}{ \delta g^{\mu\nu}(x) \delta g^{\alpha \beta}(y)} \right\rangle.
\end {align}
\end {subequations} 
These tensors are symmetric in the following way. If $\mathbf{A}$ is any of the above kernels than
\begin {equation}
\label{bitensor}
A_{\mu\nu\alpha\beta} = A_{\nu\mu\alpha\beta} = A_{\mu\nu\beta\alpha}
\end {equation}
In Eq.~\eqref{response}, $\mathbf{H}$ is manifestly causal and
$\mathbf{H}^{S,A}$ are symmetric/anti-symmetric parts of $\mathbf{H}$ respectively,
\begin {equation}
H^{S,A}_{\mu\nu\alpha\beta}(x,y) = \pm H^{S,A}_{\alpha\beta\mu\nu}(y,x).
\end {equation}
and they are
\begin {subequations}
\begin {align}
\hat{H}^S_{\mu\nu\alpha\beta} (x,y) &=   \textbf{Im} \left\langle \mathbb{T}^* \left\lbrace \hat{t}_{\mu\nu}(x) \hat{t}_{\alpha\beta}(y) \right\rbrace\right\rangle,\\
\hat{H}^A_{\mu\nu\alpha\beta} (x,y) &=   \frac{-i}{2} \left\langle  \left[ \hat{t}_{\mu\nu}(x), \hat{t}_{\alpha\beta}(y) \right]\right\rangle.
\end {align}
\end {subequations}

As we show in the Appendix~\ref{app:Kernels}, the second order variation is a matrix in time-path space is
\begin {multline}
\label{second_order}
 \frac{\delta^2 \Gamma_{ph}}{\delta \vec{\mathbf{g}} \delta \vec{\mathbf{g}}} = \frac{1}{4}
\begin{pmatrix}
i \mathbf{N} - \mathbf{H}^S - \mathbf{K}  & -i\mathbf{N} - \mathbf{H}^A \\
-i \mathbf{N} + \mathbf{H}^A & i\mathbf{N} + \mathbf{H}^S + \mathbf{K}
\end{pmatrix}.
\end {multline}

All of these kernels are real.  $\textbf{H}^A$, the anti-symmetric component of the causal response function $\textbf{H}$, changes sign in Eq.~\eqref{second_order} when the forward/backward branches are switched. Therefore it breaks time-reversal symmetry and therefore gives rise to dissipative effects. 
\subsection{Global Thermal Equilibrium and Fluctuation Dissipation Relation}

Here we show, at global thermal equilibrium, that the noise kernel $\mathbf{N}$ and the dissipation kernel $\mathbf{H}$ are related in a stationary flow. For example, a flow through a tube with varying cross section, but constant in time for each point in the tube and the temperature measured in the lab is uniform throughout the system.

The stationary flow with global thermal equilibrium is 
especially important because in the analogue gravity language it 
maps on to a hot curved space-time at global thermal 
equilibrium. Such a space contains globally time-like Killing 
vectors $\kappa^\mu$.~
\cite{birrell1984quantum,DeWitt} The local temperature in the space-time various according to the norm of the time-like Killing vector and is dictated by Tolman's law:
\begin {equation}
\label{Tolman}
T_{Tol} \sqrt{g_{\mu\nu} \kappa^{\mu} \kappa^\nu} = constant.
\end {equation}
Once there are globally time-like Killing vectors, modes that solve the covariant wave equation Eq.~\eqref{cov_wave} are classified as positive/negative Killing frequencies according to the value of their directional (Lie) derivative along the Killing vector, that is
\begin {subequations}
\begin {align}
\mathfrak{L}_\kappa u_n &= -i \omega_n u_n,\\
\mathfrak{L}_\kappa u^*_n &= i \omega_i u^*_n, \quad \omega > 0.
\end {align}
\end {subequations}
Now the field operator is written in the second quantization language and reads
\begin {equation}
\hat{\phi} = \sum_{n} \hat{a}_n^\dagger u^*_n + \hat{a}_n u_n,
\end {equation}
where the vacuum state is uniquely defined. The Hamiltonian operator with eigenvalues equal to the Killing frequencies can be written in terms of the positive energy mode operators
\begin {equation}
\label{hamiltonian}
\hat{H} = \sum_n \omega_n \hat{a}_n^\dagger \hat{a}_n .
\end {equation}
As long as the conserved energy $\omega$ and the associated temperature is used, the methods of thermal field theory in flat space-time is easily generalized to the curved space-time case.~\cite{Martin_Verdaguer_semiclassical_gravity} In the fluid system, this is not a surprise, as in a stationary case, the Hamiltonian associated with the Lagrangian in Eq.~\eqref{matter} is time independent and second quantized in the form of Eq.~\eqref{hamiltonian} .

In analogue gravity systems, this energy $\omega$ coincides with the lab frame energy of the mode.~\cite{Parentani}  Note that, this is not the comoving frame energy or the energy measured by observers in curved space-time, which is not conserved. This means the temperature in the comoving frame and the lab frame and the different and are related to each other through the Tolman law.~\cite{Volovik_book} 
Stationary flow means the Killing vector is $\kappa = (1, 0 , 0 , 0)$ In global thermodynamic equilibrium, the positive energe modes $u_i$ are distributed thermally according to the lab frame temperature so that we have

\begin {equation}
T_{Tol} \sqrt{g_{00}} = T_{Tol} \sqrt{\rho c (1 - v^2/c^2)} = T_{lab} = \frac{1}{\beta}.
\end {equation}
 Note that this analogue Tolman law is a purely classical effect that stems from the emergent Lorentz invariance of excitations in the strongly-correlated superfluid. It is to be distinguished from the Unruh effect, where an accelerating observer detects a thermal radiation of phonons even if the lab frame's temperature is zero.

In equilibrium the eigenstates are distributed according to the lab frame temperature, with the density operator of defined in Eq.~\eqref{phonon_density_operator} that reads
\begin {equation}
\boldsymbol{\hat{\varrho}}_{\phi} = \exp(-\beta \hat{H})/\text{Tr}\left\lbrace\boldsymbol{\hat{\varrho}}_{\phi}\right\rbrace.
\end {equation}
Then, the operator averages are best handled in imaginary time ($i t = \tau$) formalism. For example for two operators $\hat{O}_1$ and $\hat{O}_2$ we have, suppressing the space indices,

\begin {multline}
\label{matsubara}
\left\langle \hat{O}_1(\tau) \hat{O}_2(\tau') \right\rangle = \text{Tr}\left(  \hat{O}_1(\tau) \hat{O}_2(\tau') \hat{\rho} \right) \\= \text{Tr}\left( U(i\tau) \hat{O}_1 U(-i \tau) U(i \tau') \hat{O}_2 U(-i\tau') U(-i\beta) \right)\\
= \text{Tr}\left(  \hat{O}_2(\tau' + \beta) \hat{O}_1(\tau) \hat{\rho} \right) = \left\langle \hat{O}_2(\tau' + \beta) \hat{O}_1(\tau) \right\rangle.
\end {multline}
Stationarity allows the use of Fourier transformation  in time domain, hence this equation can be written as
\begin {equation}
\label{matsubara_w}
\left\langle \hat{O}_1 (\omega)\hat{O}_2(-\omega) \right\rangle = e^{\beta \omega} \left\langle \hat{O}_2(-\omega) \hat{O}_1(\omega) \right\rangle.
\end {equation}

Now we show that a fluctuation dissipation relation exists. Inspection of 
Eq.~\eqref{Keldysh_Langevin},~\eqref{second_order} and \eqref{noise} reveals that  the time-
reversal odd piece $\mathbf{H}^A$ of the causal response causes dissipation and the 
fluctuation kernel $\mathbf{N}$ causes noise. 

Defining
\begin {equation}
F_{\mu\nu\alpha\beta}(\omega, \vec{x},\vec{y}) = \int \frac{\text{d}\omega}{2\pi} e^{i (t-t')} \left\langle  \hat{t}_{\mu\nu}(t,\vec{x}) \hat{t}_{\alpha\beta}(t',\vec{y}) \right\rangle,
\end {equation}
and rewriting  the definitions in Eq.~\eqref{kernels} in frequency domain and using Eq.~\eqref{matsubara_w}  we get:

\begin {subequations}
\begin {align}
\hat{H}^A_{\mu\nu\alpha\beta} (\omega,\vec{x},\vec{y}) &=  -\frac{i}{2}F(\omega, \vec{x},\vec{y})\left(1 - e^{-\beta \omega}\right), \\
\hat{N}_{\mu\nu\alpha\beta} (\omega,\vec{x},\vec{y}) &=  \frac{1}{2}F(\omega, \vec{x},\vec{y})\left(1 + e^{-\beta \omega}\right).
\end {align}
\end {subequations} 
The fluctuation dissipation relation immediately follows as:
\begin {equation}
\label{FDT}
\hat{H}^A_{\mu\nu\alpha\beta} (\omega,\vec{x},\vec{y}) =  -i \tanh(\beta \omega / 2 ) \hat{N}_{\mu\nu\alpha\beta}  (\omega,\vec{x},\vec{y}).
\end {equation}

\subsection{Hydrodynamic fluctuations around a deterministic solution }
Suppose $\bar{\rho}$ and $\bar{\theta}$ are solutions to the analogue Einstein equations Eq.~\eqref{euler}. Define the classical and quantum corrections to the solutions 
\begin {subequations}
\begin {align}
\rho^\pm &= \bar{\rho} + \rho^\pm =\bar{\rho} + \rho^c \pm \rho^q\\
\theta^\pm &=\bar{\theta} + \theta^\pm =\bar{\theta} + \theta^c \pm \theta^q.
\end {align}
\end {subequations}
Let the corresponding perturbations to the metric tensor be $h^c + h^q$ and $h^c- h^q$ so that 
\begin {subequations}
\begin {align}
\vec{\mathbf{g}} &= (\bar{g}^{\mu\nu} +(h^q) ^{\mu\nu}+ (h^q) ^{\mu\nu}, \bar{g}^{\mu\nu} + (h^c)^{\mu\nu} - (h^q) ^{\mu\nu} ).
\end {align}
\end {subequations}

Breaking the deviations into classical and 
quantum parts amounts to (Keldysh) rotating the matrix in Eq.~\eqref{second_order}. The classical 
deviation can be thought of as a real displacement while the quantum deviation as a virtual displacement. 
Variation with respect to virtual displacement ($\mathbf{h}^q$) gives the classical equations of motion 
with one loop corrections as in Eq.~\eqref{euler}. To find the fluctuations of the stress-energy, we can 
expand the effective actions as a Taylor series in the classical and quantum deviations to second order 
as 

\begin {equation}
\label{Keldysh_Langevin}
\Gamma_{ph} [\vec{\mathbf{g}}]=  \mathbf{T} \cdot \mathbf{h}^q - \frac{1}{2} \: \mathbf{h}^q\cdot (\mathbf{H} + \mathbf{K}) \cdot \mathbf{h}^c +  \frac{i}{2}\: \mathbf{h}^q \cdot \mathbf{N} \cdot \mathbf{h}^q + O(\mathbf{h}^3),
\end {equation}
where, all the kernels are functionals of the unperturbed metric $g$ and  $\vec{\mathbf{h}}$ is decomposed as $(\mathbf{h}^c,\mathbf{h}^q)$ and the dot product on the right hand side contracts space-time indices with $g$ and integrates over space-time. We notice that the equation~\eqref{Keldysh_Langevin} is in the same form as the Keldysh action for a quantum particle in a bath\cite{Altland_book}. The well established technique to derive Langevin type equation from this action is to define an auxiliary stochastic tensor $\xi$ to decouple the $O((h^q)^2)$ term (the Hubbard-Stratanovich decomposition). 

\begin {multline}
\label{Hubbard_Strat}
\exp\left(i\Gamma_{ph} [\vec{\mathbf{g}}] \right)\\
 = \frac{1}{\sqrt{\text{Det}(2\pi \mathbf{N})}} \int \mathscr{D}[\boldsymbol{\xi}]e^{-\frac{1}{2} \xi \cdot \mathbf{N}^{-1} \cdot \xi}e^{i \mathbf{Re} \lbrace\Gamma_{ph}\rbrace + i \boldsymbol{\xi}\cdot \mathbf{h}^q}\\
= \left\langle  e^{i \mathbf{Re} \lbrace\Gamma_{ph}\rbrace + i \boldsymbol{\xi}\cdot \mathbf{h}^q} \right\rangle_{\xi} = \left\langle  \exp\left(i \Gamma_{ph}[\vec{\mathbf{g}}; \boldsymbol{\xi}]\right) \right\rangle_{\xi}.
\end {multline}

The mean and correlation function of the noise tensor field $\xi$ is defined by the above Gaussian path integral, using the definition Eq.~\eqref{noise} of the noise kernel $\mathbf{N}$,  which yields
\begin {subequations}
\label{noise}
\begin {align}
\left\langle \xi_{\mu\nu}(x) \right\rangle_{\xi} &= 0,\\
\left\langle \xi_{\mu\nu}(x) \xi_{\alpha\beta}(y) \right\rangle_{\xi} &= \frac{1}{2}\left \langle \lbrace \hat{t}_{\mu \nu}(x), \hat{t}_{\alpha \beta}(y)\rbrace\right\rangle.
\end {align}
\end {subequations}

The new phonon effective action 
\begin {equation}
\label{Keldysh_Langevin}
\Gamma_{ph} [\vec{\mathbf{g}}; \boldsymbol{\xi}[\bar{g}]]=  \left(\mathbf{T} + \boldsymbol{\xi}\right) \cdot \mathbf{h}^q - \frac{1}{2} \: \mathbf{h}^q\cdot (\mathbf{H} + \mathbf{K}) \cdot \mathbf{h}^c + O(\mathbf{h}^3)
\end {equation}
depends on the noise tensor $\xi$.

Then the background plus phonon effective action in Eq.~\eqref{first_loop} as a functional of noise tensor 
\begin {equation}
\Gamma[\rho^\pm, \theta^\pm , \boldsymbol{\xi}] =  S[\rho^+ , \theta^+] - S[\rho^- , \theta^-] + \Gamma_{ph}[\vec{\mathbf{g}}; \boldsymbol{\xi}[\bar{g}]].
\end {equation}

\subsection{Stochastic analogue Einstein equation}
The Euler-Lagrange equations for the stochastic effective action $\Gamma[\rho^\pm, \theta^\pm , \boldsymbol{\xi}] $ in the limit $(\rho^\pm ,\theta^\pm) \to (\rho ,\theta)$ and hence $h^q \to 0$ are the following stochastic analogue Einstein equations for the two-fluid hydrodynamics

\begin {subequations}
\label{noisy_euler}
\begin {align}
\!\!&\partial_t \rho + \nabla \cdot(\rho \vec{v}) = \frac{1}{2} \nabla \cdot \left(\sqrt{-g} \left[\left\langle \hat{T}_{\mu\nu} \right\rangle + \xi_{\mu\nu}\right] \frac{\delta g^{\mu\nu}}{\delta \vec{v}}\right);\\
\!\!&\partial_t \theta + \frac{1}{2} v^2 + \mu(\rho)  = \frac{1}{2} \left(\sqrt{-g} \left[\left\langle \hat{T}_{\mu\nu} \right\rangle  + \xi_{\mu\nu}\right]  \frac{\delta g^{\mu\nu}}{\delta \rho}\right).
\end {align}
\end {subequations}
In this equation, the  noise source $\boldsymbol{\xi}[\bar{g}]$ is computed by using the metric $\bar{g}
(\bar{\rho},\bar{{v}})$ that solves Eq.~\eqref{euler}, while $\rho$ and $\theta$  suffer  deviations 
$\rho^c, \theta^c$ from the solution of Eq.~\eqref{euler} due to the existence of noise. The energy 
momentum tensor is also computed by taking these deviations in to account. Next we summarize the procedure to find the stochastic deviations  $\rho^c$ and $\theta^c$ of the background.

\subsection {Procedure to compute background metric fluctuations from the analogue Einstein equation}
\label{sec:procedure}

Below, we outline the procedure to determine  fluctuations of the metric, given initial data that consists of the initial conditions, boundary conditions,  an initial state of the phonon density operator, and external sources, if any.

First, is to compute the deterministic solution $(\bar{\rho},\bar{{v}})$ and the associated metric $\bar{g}_{\mu\nu}[\bar{\rho},\bar{\theta}]$ as defined in Eq.~\eqref{metric}, by solving the analogue Einstein equation~\eqref{euler} \textit{self consistently}. In general this is  equivalent to solving the two-fluid equations Eq.~\eqref{two-fluid} and is a difficult task. However, one can approach this problem perturbatively.  If one has a solution to the Euler equations that describes the background without the phonons (Eq.~\eqref{GPE_hydro} with no quantum pressure), then the expectation value of the stress-energy operator Eq.~\eqref{stress_renorm} is computed by using the techniques in Ref.~[\citen{birrell1984quantum}] and references therein. One then continues this process up to a  desired order, so that the background is consistent with the metric on which the sources are computed. 

Second, is to compute the correlators of the stress tensor in Eq.~\eqref{kernels} to form the response and noise kernels. For the methods required to perform this, we refer the reader to Ref.[~\citen{Noise_kernels},~\citen{kernels2}] and references therein.      

Third is to linearize the stochastic analogue Einstein equation~\eqref{noisy_euler} around the deterministic solution $\bar{\rho}, \bar{\theta}$. The resulting equation is linear in the stochastic corrections  $\rho^c$ and $\theta^c$,  to the background variables. The associated first-order corrections to the deterministic metric $\bar{g}$ follow from the chain rule as
\begin {equation}
\label{hc}
(h^c)^{\mu\nu} = \rho^c  \frac{\delta \bar{g}^{\mu\nu}}{\delta {\rho}}  + \vec{v}^c \cdot \frac{\delta \bar{g}^{\mu\nu}}{\delta \vec{v}}.
\end {equation}
Then, the linearized stochastic analogue Einstein equation, that we dub the analogue Einstein-Langevin equation, follows from Eq.~\eqref{noisy_euler}, where we linearize the stress tensor   by using Eq.~\eqref{Keldysh_Langevin}, as
\begin {multline}
\label{lin1}
\partial_t \rho^c + \nabla \cdot(\bar{\rho} \vec{v}^c) + \nabla \cdot(\rho^c \bar{{v}})=  \frac{1}{2} \nabla \cdot \bigg(\sqrt{-\bar{g}} \frac{\delta \bar{g}^{\mu\nu}}{\delta \bar{v}}  \\\times\left[-\frac{1}{2} \int \text{d}^4 y (H+K)_{\mu\nu\alpha\beta}(x,y) (h^c)^{\alpha\beta}(y)+ \xi_{\mu\nu}\right] \bigg)
\end {multline}
and
\begin {multline}
\label{lin2}
\partial_t \theta^c + \bar{{v}}\cdot \vec{v}^c + \frac{\bar{c}^2}{\bar{\rho}}\rho^c=  \frac{1}{2} \bigg(\sqrt{-\bar{g}} \frac{\delta \bar{g}^{\mu\nu}}{\delta {\rho}}  \\\times\left[-\frac{1}{2} \int \text{d}^4 y (H+K)_{\mu\nu\alpha\beta}(x,y) (h^c)^{\alpha\beta}(y)+ \xi_{\mu\nu}\right] \bigg).
\end {multline}

Fourth and final,  is to write down  correlators of $h^c$. To do this, one first expresses the correlators of the background corrections $\rho^c$ and $\theta_c$ in terms of the noise correlator $\mathbf{N}$ by using the Green's functions of the linear equations Eq.~\eqref{lin1} and ~Eq.~\eqref{lin2}. Then the correlators of $h^c$ follow from Eq.~\eqref{hc}. 

\section{Equilibrium `Minkowski' fluctuations}
\label{minkowski}

In this section we illustrate the procedure of Sec.~\ref{sec:procedure} for the static ($\vec{v}_s = 0$) thermodynamic equilibrium focusing on a qualitative analysis (detailed technical calculations and results for the correlators are cumbersome and will be presented elsewhere).  The metric tensor for the static background reduces to that of Minkowski, if one sets
\begin {equation}
\bar{\rho} = \rho_e = c_e = \bar{c} = 1
\end {equation}
in addition to the choice of units in Eq.~\eqref{units}.

This renders Step 1 in Sec.~\ref{sec:procedure} trivial. The analogue Einstein equations Eq.~\eqref{euler} and the covariant conservation law Eq.~\eqref{cov_con} are trivially satisfied.
Nevertheless, it is instructive to outline how the expectation value of the stress-tensor is computed. 

The isotropy and homogeneity of Minkowski space places strong restrictions on the stress-energy tensor and its correlators.
For  the stress tensor, homogeneity means all components are constant in space-time. The trace of the expectation value of stress energy vanishes for a massless scalar field. The energy density is positive. There is no isotropic Cartesian vector, therefore the momentum density vanishes. 
The only isotropic rank-2 Cartesian tensor is the Kronecker delta, therefore the spatial components are isotropic.  The only rank-2  Minkowski tensor with the above properties is

\begin {equation}
\left\langle \tensor{\hat{T}}{^\mu}{_\nu}\right \rangle = 
 \frac{\pi^2 T^4}{90}
\begin{pmatrix}
3 & 0 & 0 & 0 \\
0 & -1 & 0 & 0 \\
0 & 0 & -1 & 0 \\
0 & 0 & 0 & -1
\end{pmatrix},
\end {equation}
where the overall factor is the black body energy density and can be obtained from the standard momentum space integral that arise in Eq.~\eqref{stress_renorm}, after dropping the infinite vacuum energy.

Similarly to the stress tensor, the symmetries greatly reduce the complexity of calculating its correlates as well. First of all, the homogeneity of space-time requires that any function $f(x,y) = f(x-y)$. The stress tensor correllators can be expressed as momentum space integrals. In thermal equilibrium, all non-local behavior is due to the factor $\coth(\beta  \omega / 2)$, which behaves like
\begin {equation}
\label{locality}
 \coth(\beta \omega/2) \xrightarrow{k_B T \gg \hbar \omega}\frac{2 k_B T}{\hbar \omega} 
\end {equation}
in the high-temperature limit and all kernels become local (memoryless) operators in space-time.

In this limit, we linearize Eq.~\eqref{noisy_euler} around $\bar{\rho}$ and $\bar{\theta}$ to get Eq.~\eqref{lin1} and ~\eqref{lin2} for the static equilibrium. This linearized equation is in the Langevin form. In Fourier domain it reads
\begin {equation}
\label{Langevin_wave}
\!\!\!\!\begin{pmatrix}
-i\omega & -|\vec{k}|^2\\
1 & -i\omega
\end{pmatrix}
\begin{pmatrix}
\rho^c \\ \theta^c
\end{pmatrix}
=
\begin{pmatrix}
R_{11} & R_{12}\\
R_{21} & R_{22}
\end{pmatrix}
\begin{pmatrix}
\rho^c \\ {\theta}^c
\end{pmatrix}
+
\begin{pmatrix}
\zeta_1 \\ \zeta_2
\end{pmatrix}.
\end {equation}
where $\hat{R}$'s are local operators, hence polynomials in $\vec{k}$ and $\omega$ and $\zeta$'s are stochastic sources.

The stochastic sources $\zeta$'s are deduced from $\xi_{\mu\nu}$ by applying Eq.~\eqref{contract}, as
\begin {subequations}
\begin {align}
\zeta_1 &= -i k^j \xi_{0 j}\\
\zeta_2 &= \frac{1}{2} (\kappa - 1) \tensor{\xi}{^\mu_\mu} -\kappa \xi^{00}
\end {align}
\end {subequations}
where $\kappa = \partial \ln c / \partial \ln \rho$ is the change of energy of phonons with density. 

The diagonal terms $R_{11}$ and $R_{22}$ in Eq.~\eqref{Langevin_wave} are equal and the symmetric part of the response kernel does not contribute to them:
\begin {equation}
 {R}_{11} =  {R}_{22} = -i\frac{(1 -\kappa)}{2} k^j \tensor{(H^A)}{^0_j^\mu_\mu}  \\-i {\kappa}  k^j\tensor{(H^A)}{^0_j^0^0}.
\end {equation}
Whereas, the antisymmetric part of the response kernel does not contribute to the off diagonal components
\begin {equation}
 {R}_{12}  = k^i k^j \tensor{(H^S+K)}{^0_i^0_j} 
\end {equation}
and
\begin {multline}
 {R}_{21}  = -\frac{(1 -\kappa)^2}{4} \tensor{(H^S+K)}{^\nu_\nu^\mu_\mu}  \\ -\kappa^2 \tensor{(H^S+K)}{^0^0^0^0} +\kappa(\kappa-1) \tensor{(H^S + K)}{^\mu_\mu^0^0}.
\end {multline}

Now one can solve Eq.~\eqref{Langevin_wave} and express the correlators of stochastic corrections $\rho^c$ and $\theta^c$ in terms of the noise correlators and the metric fluctuation follows by using Eq.~\eqref{hc}.

We also note that, to lowest order in $k$, introducing some constants $\alpha, \beta, \nu $ we have
\begin {subequations}
\begin {align}
{R}_{11} =  {R}_{22} &\to -\nu |\vec{k}|^2\\
{R}_{12}  & \to \alpha_{i j } k^i k^j\\
{R}_{21} &\to -\beta
\end {align}
\end {subequations}
and the Eq.~\eqref{Langevin_wave} takes the form of the linearized Navier-Stokes and continuity equations with stochastic forcing terms

\begin {subequations}
\label{Stochastic_NS}
\begin {align}
&\partial_t \rho^c + (\delta_{i j} + \alpha_{i j}) \nabla_i \vec{v}^c_j - \nu \nabla^2 \rho^c = \zeta_1, \\
&\partial_t \vec{v}^c - \nu\nabla^2 \vec{v}^c =
-(1+ \beta) \nabla \rho^c +{\nabla}\zeta_2.
\end {align}
\end {subequations}
The coefficient of viscosity $\nu$ can be extracted as
\begin {equation}
\label{viscosity}
\nu = \frac{1}{3}\frac{\partial}{\partial {k_j}}\left[\frac{(1 -\kappa)}{2} \tensor{(H^A)}{^0_j^\mu_\mu}  - {\kappa}  \tensor{(H^A)}{^0_j^0^0}\right]\bigg\vert_{\vec{k}=0} = 0.
\end {equation}
The vanishing viscosity is because the components of the anti-symmetric response tensor $H^A$ that contribute to $\nu$ are zero by virtue of the fluctuation dissipation relation in Eq~\eqref{FDT} and the noise tensor $N$ for flat spacetime that is readily computed in Ref~[\citen{flat_kernels}]. This is in agreement with the fundamental premise of the two-fluid model, namely that the mutual viscosity between the normal and superfluid components is zero.

Ignoring $\alpha_{i i } \ll 1$ and $\beta \ll 1$, since the operating temperatures are small compared to internal energy of fluid and dropping off-diagonal components of $\alpha$ due to consideration of isotropy Eq.~\eqref{Stochastic_NS} reduces to:

\begin {subequations}
\begin {align}
&\partial_t \rho^c + \nabla^2 {\theta}^c  = \zeta_1, \\
&\partial_t {\theta}^c  +
  \rho^c =\zeta_2.
\end {align}
\end {subequations}
 Putting $\kappa = 0$ for simplicity and defining the (retarded) Green's function
\begin {equation}
\label{Green}
\mathbf{G}=  
\frac{1}{\omega^2 - |\vec{k}|^2}
\begin{pmatrix}
|\vec{k}|^2 & \omega k^j\\
 -i\omega & -i k^j
\end{pmatrix},
\end {equation}
the solution is 
\begin {equation}
\label{Langevin_wave}
\begin{pmatrix}
\rho^c \\ {\theta}^{c}
\end{pmatrix}
=
\begin{pmatrix}
G_{11} & G_{12}\\
{G}_{21} & {G}_{22}
\end{pmatrix}
\begin{pmatrix}
\frac{1}{2}\tensor{\xi}{^\mu_\mu} \\ {\xi}_{0 j}
\end{pmatrix}.
\end {equation}

The Green's function in Eq.~\eqref{Green} relates the covariant noise tensor $\xi$ to the density and noise fluctuations  $\rho^c$ and $\vec{v}^c$.
To streamline the notation we define
\begin {subequations}
\begin {align}
\tensor{\tilde{N}}{^\mu_\mu^\nu_\nu} &= 8\pi^4 T^8 n= \frac{\pi^4 T^8}{1350}, \\
 \tensor{\tilde{N}}{_0_1_0_1} &= \pi^4 T^8 n = \frac{1}{4}\left\langle \tensor{\hat{T}}{_{0}_{0}}\right\rangle \left\langle\tensor{\hat{T}}{_{1}_{1}}\right\rangle = \frac{\pi^4 T^8}{10800} ,
\end {align}
\end {subequations}
where the tensor $\tilde{N}$ is the momentum space representation of the local noise kernel. The values of $N(x-y)$ at the coincidence limit is readily computed in Ref~[\citen{flat_kernels}].

The correlation function of fluctuations reads, for example,
\begin {multline}
\label{rho_v_F}
\left\langle \rho^c(k)\theta^c(-k)\right\rangle = n \pi^4 T^8 \\ \times
(2   G_{11}(k) {G}_{21}(-k) +  G_{12}(k) {G}_{22}(-k)) ,
\end {multline}
The expectation value $\langle. \rangle$ is computed with respected to the noise $\xi$, as in Eq.~\eqref{Hubbard_Strat}. The correlation function Eq.~\eqref{rho_v} reads:
\begin {equation}
\label{rho_v}
\left\langle \rho^c(\omega,\vec{k}) \vec{v}^c(-\omega,-\vec{k})\right\rangle =
3n  \pi^3 T^8 \frac{\omega\vec{k} |\vec{k}|^2}{(\omega^2-|\vec{k}|^2)^2}.
\end {equation}

We obtain the other non-zero correlation functions in a similar fashion as  
\begin {equation}
\label{rho_rho}
\left\langle \rho^c(\omega, \vec{k}) \rho^c(-\omega, -\vec{k})\right\rangle = n \pi^3 T^8 \frac{|\vec{k}|^2 (\omega^2 + 2 |\vec{k}|^2)}{(\omega^2-|\vec{k}|^2)^2},
\end {equation}
and finally
\begin {equation}
\label{v_v}
\left\langle \vec{v}^{c i}(\omega, \vec{k}) \vec{v}^{c j}(-\omega, -\vec{k})\right\rangle = n \pi^3 T^8 k^i k^j \frac{2 \omega^2 + |\vec{k}|^2}{(\omega^2-|\vec{k}|^2)^2}
\end {equation}

Here the denominators are regularized so that there are equal number of poles in the upper and lower complex planes. Note that, the denominator $\omega^2 -|\vec{k}|^2$ implies that the stochastic corrections to density and velocities between two points are correlated as long as the two points are sound-like separated (i.e. $t^2 - |\vec{x}|^2 = 0$) and therefore the correlation functions are non-local. This has the following physical interpretation. When the phonon field creates a disturbance on ,say, the density of the fluid at a point, the disturbance emanates with the speed of sound and effects the noise corrections to density at all points where the sound can reach. The forms of these correlators are reminiscent of the correlators of relativistic hydrodynamics [\citen{Kovtun}]. However, the $T^8$ dependence is specific to the two-fluid model, arising due to the fact that noise tensor can be written as a product of two phonon stresses, each of which scale with $T^4$.  

\section{Stochastic Lensing of acoustic waves: a conjecture}

The main fundamental physical consequence of our analysis is the finding that sound waves propagating on a superfluid background at a finite temperature experience a random, stochastic metric on top a dynamical but deterministic  background metric determined by the classical superfluid flow.
For example, when the deterministic background metric is Minkoswki i.e. the superfluid is in static equilibrium, the stcohastic metric reads

\begin {equation}
g^{\mu\nu} = \eta^{\mu\nu} + (h^c)^{\mu\nu}
\end {equation}

where $\eta$ is the Minkoswki metric and $h^c$ is computed using Eq.~\eqref{hc} by evaluating the derivatives of the metric at Minkowski. The resulting stochastic component reads

\begin {equation}
(h^c)^{\mu\nu} = 
\begin{pmatrix}
-(1+\kappa)\rho^c & \vec{v}^{cT}\\
\vec{v}^c & (1-\kappa)\rho^c \mathbf{I}_{3\times 3} 
\end{pmatrix}.
\end {equation} 
Here, $\rho^c$ and $\vec{v}^c$ are solutions to the stochastic Navier-Stokes equation Eq.~\eqref{Langevin_wave} or~\eqref{Stochastic_NS}. Similarly we define the two point correlation function for the metric fluctuations as the following symmetric tensor
\begin {equation}
M^{\mu\nu\alpha\beta}(\vec{r},t) = \left\langle(h^c)^{\mu\nu}(\vec{r},t)(h^c)^{\alpha\beta}(0)\right\rangle.
\end {equation}
The components of this tensor follows from Eqs.~\eqref{rho_v},~\eqref{rho_rho} and ~\eqref{v_v}. For example for $\kappa = 0$
\begin {equation}
M^{0  0 0 0}(\vec{r},t) = -M^{0  0 i i }(\vec{r},t) = M^{i i  j j}(\vec{r},t)= \left\langle\rho^c(\vec{r},t)\rho^c(0)\right\rangle.
\end {equation}
The other non-zero components of $M$ up to symmetries are $M^{0i0i}$ and $M^{000i}$ that are $\langle v^i(\vec{r},t) v^i_2(0) \rangle$ and $\langle \rho(\vec{r},t) v^i_2(0) \rangle$ respectively. 

 This means, even in the absence of any flows and in equilibrium, where the average metric is that of flat Minkowski space, the acoustic ``rays'' would experience random deviations from the flat background. This may lead to analogue gravitational lensing of acoustic waves due to the phenomenon called `intermittency'. In Ref. [\citen{Zeldovich}], Zeldovich considered light rays propagating in a random medium with a fluctuating metric, and showed that even if the average metric is flat, an observer receiving two distant rays would see the rays bend and the corresponding object shrink, due to the stochasticity. The effect seems unobservable in actual general relativity due to very weak metric fluctuations, if any.\cite{Sokoloff}  However, similar fluctuations of ``synthetic metric'' can be greatly enhanced in a superfluid  and it is conceivable that the corresponding analogue Zeldovich effect -- bending of acoustic``rays'' propagating through a thermal superfluid -- could become observable.

\section{Conclusions and Outlook}

In this paper, we  develop a geometric generalization of the Landau-Khalatnikov two-fluid hydrodynamics for a strongly-correlated bosonic superfluid and derive a system of coupled equations describing the motion of the superfluid background and the entropy-carrying normal fluid of phonon excitations. By exploiting the emergent covariance of the phonon field, we draw analogies between general relativity and the two-fluid system  including  stochastic effects due to the fluctuations of the phonon ``matter'' field. 

We emphasize that the methods and analogies used here are applicable to a variety of systems. Indeed, weakly interacting quasiparticles excited over a ground state is a common picture in many-body physics. Hydrodynamics captures the long-wavelength behavior of such systems. Galilean invariance requires that quasiparticle dynamics is defined on the frame comoving with the background fluid that represents the ground state manifold. This means, when referred to the lab frame, that the quasiparticle field experiences shifts due to the background flow -- an effect captured in the definition of the metric tensor~(\ref{metric}). From thereon, the quasiparticles can be described by quantum field theory on curved space-time, thereby justifying the geometric formulation of the problem. 

We would like to conclude with a brief summary of our work and promising directions for further research

In Sec.~\ref{sec:vacuum}, following an earlier work~\cite{Volovik_book}, we discussed the applicability of the metric description for fluid by analyzing the length- and energy-scales. In analogy with cosmology~\cite{Volovik_cosmology, Volovik_superfluid}, the geometric description breaks down at an effective ``Planck energy,'' where  the hydrodynamic description becomes inapplicable. However, the quantum description of the boson system is known at all energies contrary to the case in cosmology. This paves the way to understanding the cosmological phenomena through condensed matter systems.~\cite{Volovik_qbrane,Volovik_Higgs,Volovik_weyl}

In Sec.~\ref{sec:matter}, we used the background field formalism to write down the Keldysh theory of quasiparticles. We use the Keldysh contour because  the dynamical fluctuations of the phonon field follow naturally from the Keldysh effective action that is ``designed'' to account for non-equilibrium phenomena. One interesting finding of this analysis is an additional ``gravitational  anomaly'' , not present in the original work of Unruh~\cite{Unruh} and Stone~\cite{Stone} : that is, despite the action for the phonons is covariant, as already noted in the literature, the measure of the path integral is not covariant. Hence, in contrast to classical phonon action, their full quantum theory is not covariant. We believe this anomaly should be taken into account, whenever quantum effects such as Hawking radiation, is important. It is curious to see if this anomaly is related to gravitational anomalies in high-energy phenomena. These interesting issues will be discussed in a  future publication. 

In Sec.~\ref{sec:two-fluid},  inspired by Ref.~\citen{Volovik_book}, we derive the analogue Einstein equation that governs the background and the excitations (matter). We establish the equivalence of the analogue Einstein equation and the covariant conservation law for the phonons, to the  two-fluid conservation laws. Instead of assuming a specific form for the stress-energy tensor as in Ref.~[\citen{Volovik_cosmology}], we prove the equivalence of the two descriptions by reducing the covariant conservation law  down to the Noether current of the two-fluid system and by using the equations of motion.  We believe that the proof presented here is more general, than the specific superfluid model we study, and should be applicable to any emergent general-relativistic theory, which may be derived in the hydrodynamic, long-wave-length limit of a ``parent'' condensed matter system.

 In Sec.~\ref{sec:stochastic}, we take the analogy between the superfluid system and general relativity further to the domain of stochastic fluctuations. This allows the application of a variety of methods that have been invented for  stochastic gravity~\cite{Hu_Verdaguer_primer} to condensed matter systems.We write the response and dissipation kernels in the covariant language. In global thermal equilibrium, we discuss the notion of temperature on curved space-time and show the appearance of analogue Tolman's law. We prove the fluctuation-dissipation relation for a metric with globally time-like Killing vectors  -- that is, for a flow that can  be brought to a stationary form after a  Galilean transformation.  Finally, we outline a procedure to calculate correlators of various observables, including fluctuations of the metric.

One interesting direction would be to consider the effects of higher order correlators of the stress-energy tensor and build a hierarchy of equations similar to the BBGKY hierarchy.

We linearize the stochastic analogue Einstein equation   
and obtain a Langevin-type equation for the stochastic corrections to the background. In Sec.~\ref{minkowski}, we show that the symmetries of the flow determine the structure of the Langevin equation, by considering the Minkowski case. We solve the langevin equation and find the correlation functions of stochastic fluctuations in density and velocity. The correlation functions scale with $T^8$. The density at a point is correlated with velocities at all light-like separated points, as long as the velocity is parallel to the vector connecting the two points. Similarly co-directional velocities and densities between light-like separated points are correlated. This means the background establishes correlations between distant points, hence the metric fluctuations have a non-local correlation.  

Classification of the flow induced space-times, the stress tensor and its fluctuations on the basis of the symmetry of the flow is a well defined mathematical problem. In general relativity these correspond to the Petrov classification of space-times and Segre classification of symmetric tensors.~\cite{petrov_Einstein} Together with the machinery of general relativity such as conformal transformations and general coordinate transformations, the  classification of the solutions of the two-fluid equations can be carried out in a similar spirit.

In Sec.~\ref{sec:stress} we argued that the divergent piece of the stress tensor can be  discarded by arguing that it is renormalizes into the internal energy. This contribution to the energy denoted by $E_0$ renormalizes the pressure as explicitly seen in Eq.~\eqref{pressure}. In the cosmology context this is analogous to the renormalization of the vacuum energy into the cosmological constant. Recently, this idea is further elaborated by taking the relaxation dynamics of the quantum vacuum into account.~\cite{q-theory} The noise and dissipation created by the quantum fields might contribute to the dynamic process that leads to a Minkowski steady state with small cosmological constant.

Finally, we conjecture a lensing of sound waves due to stochastic fluctuations of the background metric analogous to the intermittent behavior of geodesics in stochastic spacetime metric in astrophysical context.

\section{Acknowledgements}

V.G. is grateful to Grigorii E. Volovik, Dmitry D. Sokoloff, Gil Refael. A.C.K. is grateful to Theodore Jacobson, Bei-Lok Hu, Raman Sundrum and Gokce Basar for valuable discussions. This work was supported by US-ARO (contract
No. W911NF1310172), NSF-DMR 1613029, and Simons Foundation. Part of this work was completed at the Kavli Institute for Theoretical Physics (KITP) and the authors are grateful to KITP for hospitality and for the partial support of this research from the National Science Foundation under Grant No. NSF PHY-1125915.

\appendix
\section{Derivation of the canonical conservation law}
\label{app:conservation}
 The Chrystoffel symbols are

\begin {equation}
\label{chris_def}
\tensor{\Gamma}{^\mu_\nu_\alpha} = \frac{1}{2} g^{\mu\beta} \left( g_{\beta\nu,\alpha} + g_{\beta\alpha,\nu} - g_{\nu\alpha,\beta}  \right),
\end {equation}
where the comma notation means
$$
g_{\beta\nu,\alpha} \equiv \frac{\partial}{\partial x^\alpha}g_{\beta\nu}.
$$
For a symmetric tensor $\tensor{T}{^\mu_\theta}$, with the application of chain rule, we  have
\begin {multline}
\label{chris_1}
\tensor{T}{^\mu_\theta}\tensor{\Gamma}{^\theta_\mu_\nu} = -\frac{1}{2} \tensor{T}{_\gamma_\theta}\left(\tensor{g}{^\gamma^\mu_{,\nu}}\tensor{\delta}{^\theta_\mu} + \tensor{g}{^\gamma^\mu_{,\mu}} \tensor{\delta}{^\theta_\nu} -  \tensor{g}{^\theta^\mu_{,\mu}} \tensor{\delta}{^\gamma_\nu}\right)\\ = -\frac{1}{2} \tensor{T}{_\gamma_\theta}\left(\tensor{g}{^\gamma^\mu_{,\nu}}\tensor{\delta}{^\theta_\mu}\right) = -\frac{1}{2} \tensor{T}{_\gamma_\mu}\tensor{g}{^\gamma^\mu_{,\nu}}.
\end {multline}

Similarly, we have

\begin {multline}
\label{chris_2}
\tensor{\Gamma}{^\mu_\nu_\mu} = \frac{1}{2} g^{\mu\beta} \left( \tensor{g}{_\beta_\nu_{,\mu}} + \tensor{g}{_\beta_\mu_{,\nu}} - \tensor{g}{_\nu_\mu_{,\beta}}  \right)\\
= \frac{1}{2} g^{\mu\beta} \tensor{g}{_\beta_\mu_{,\nu}} = \frac{1}{2 g} g_{,\nu} = \partial_\nu \log \sqrt{-g}.
\end {multline}

Using Eq.~\eqref{chris_1} and Eq.~\eqref{chris_2} and relabeling dummy indices as required, we write the covariant conservation law Eq.~\eqref{cov_con} in the main text as Eq.~\eqref{cov_con2}.

Now, by using Eq.~\eqref{canon_stress} and Eq.~\eqref{second_piece}, the covariant conservation law Eq.~\eqref{cov_con2} can be recast in the Noether current form that appears in Eq.~\eqref{Noether} as follows.

\begin {multline}
\label{step1}
\sqrt{-g}\nabla_\mu \tensor{T}{^\mu_\nu} = \left(\frac{\partial \mathscr{L}_{ph} }{\partial \phi_{,\mu}} \:\phi_{,\nu}
-\mathscr{L}_{ph} \delta{^\mu}{_\nu}\right)_{,\mu} \\+ \frac{\partial \mathscr{L}_{ph}}{\partial \theta_{,\mu}}\theta_{,\mu\nu} + \frac{\partial \mathscr{L}_{ph}}{\partial \rho} \rho_{,\nu}=0.
\end {multline}

Using the Euler-Lagrange equation (classical limit of Eq.~\eqref{cont}) 
$$
\frac{\partial(\mathscr{L}_{ph} + \mathscr{L}_{cl})}{\partial \rho} = \left(\frac{\partial(\mathscr{L}_{ph} + \mathscr{L}_{cl}) }{\partial \rho_{,\mu}}\right)_{,\mu} = 0,
$$
and using the chain rule
$$
\partial_\nu \mathscr{L}_{cl} = \frac{\partial \mathscr{L}_{cl}}{\partial \rho} \partial_\nu \rho+ \frac{\partial \mathscr{L}_{cl}}{\partial \theta_{,\mu}}  \theta_{,\mu\nu},
$$
the covaraint derivative in Eq.~\eqref{step1} becomes
\begin {multline}
\label{step2}
\sqrt{-g}\nabla_\mu \tensor{T}{^\mu_\nu} = \left(\frac{\partial \mathscr{L}_{ph} }{\partial \phi_{,\mu}} \:\phi_{,\nu}
-\mathscr{L}_{ph} \delta{^\mu}{_\nu}\right)_{,\mu} \\+ \frac{\partial (\mathscr{L}_{ph}+\mathscr{L}_{cl})}{\partial \theta_{,\mu}}\theta_{,\mu\nu} -\partial_\mu  \mathscr{L}_{cl} \tensor{\delta}{^\mu_\nu}=0.
\end {multline}
Now using, the second Euler-Lagrange equation (classical limit of Eq.~\eqref{Bernouilli}), 
$$
\left(\frac{\partial(\mathscr{L}_{ph} + \mathscr{L}_{cl}) }{\partial \theta_{,\mu}}\right)_{,\mu}= \frac{\partial(\mathscr{L}_{ph} + \mathscr{L}_{cl})}{\partial \theta} = 0.
$$ 
the Eq.~\eqref{step2} reduces to
\begin {multline}
\label{step3}
\sqrt{-g}\nabla_\mu \tensor{T}{^\mu_\nu} = \left(\frac{\partial \mathscr{L}_{ph} }{\partial \phi_{,\mu}} \:\phi_{,\nu}
-(\mathscr{L}_{cl} + \mathscr{L}_{ph} )\delta{^\mu}{_\nu}\right)_{,\mu} \\+ \left(\frac{\partial (\mathscr{L}_{ph}+\mathscr{L}_{cl})}{\partial \theta_{,\mu}}\theta_{,\nu}\right)_{,\mu}=0.
\end {multline}
The total derivative on the right hand side can be written as the divergence of the Noether current Eq.~\eqref{Noether}.

\section{Noise and Response Kernels}
\label{app:Kernels}

Here, we write the first and second order variations as tensors and bi-tensor kernels. For example consider the second order variations of the phonon effective action with respect to the forward metric:
\begin {multline}
\label{plus_plus_0}
\frac{\delta^2 \Gamma_{ph}}{ \delta (g^+)^{\mu\nu}(x) \delta (g^+)^{\alpha \beta}(y)}\bigg\rvert_{g^+ = g^-} = \\ \frac{-i}{4 } \sqrt{-g(x)} \left\langle \hat{T}_{\mu\nu}(x)\right\rangle\sqrt{-g(y)} \left\langle \hat{T}_{\mu\nu}(y)\right\rangle \\
 - i \frac{\delta^2 Z_{ph}}{ \delta (g^+)^{\mu\nu}(x) \delta (g^+)^{\alpha \beta}(y)}\bigg\rvert_{g^+ = g^-}.
\end {multline}
In the last part we used the unitarity of the partition function in Eq.~\eqref{unitarity}.

The second order variation of the phonon partition function
contains the expectation value of a product of stress-energy operators.
\begin {multline}
\label{second_order_app}
\frac{\delta^2 Z_{ph}}{ \delta (g^+)^{\mu\nu}(x) \delta (g^+)^{\alpha \beta}(y)}\bigg\rvert_{g^+ = g^-} =\\ -\frac{1}{4} \sqrt{-g(x)} \sqrt{-g(y)} \left\langle \mathbb{T}^* \left\lbrace \hat{T}_{\mu\nu}(x) \hat{T}_{\alpha\beta}(y) \right\rbrace\right\rangle\\
+ i \left\langle \frac{\delta^2 S_{ph}[\hat{\phi}]}{ \delta g^{\mu\nu}(x) \delta g^{\alpha \beta}(y)} \right\rangle.
\end {multline}
Here, the Weyl or $\mathbb{T}^*$ ordered  average of the product of stress operators is understood in the following way. The stress-energy operator contains terms quadratic in the field and its derivatives. It can be regularized by using the point splitting method.\cite{Martin_Verdaguer_stochastic_gravity,birrell1984quantum}
In this regularization scheme, to take an average as in Eq.~\eqref{second_order_app}, one first writes the quadratic operator average as
\begin {multline}
\label{Weyl_order}
 \lim_{x',x''\to x} \left\langle \mathbb{T}^* \left\lbrace \partial_\mu \hat{\phi}(x') \partial_\nu \hat{\phi}(x'') \right\rbrace\right\rangle \\=  \lim_{x',x''\to x} \frac{\partial}{\partial x^{'\mu}}  \frac{\partial}{\partial x^{''\nu}} \left\langle \mathbb{T} \left\lbrace  \hat{\phi}(x') \hat{\phi}(x'') \right\rbrace\right\rangle.
\end {multline} 
Then, the time ordered operator average is computed and renormalized by subtracting the divergent terms. Lastly, the operators are acted on the result and the coincidence limit is taken.  

Weyl ordering inherits all properties of time or path ordering, for example it is possible to rewrite the Weyl ordered average as
\begin {multline}
 2 \left\langle \mathbb{T}^* \left\lbrace \hat{T}_{\mu\nu}(x) \hat{T}_{\alpha\beta}(y) \right\rbrace\right\rangle =   \left\langle \left\lbrace \hat{T}_{\mu\nu}(x),\hat{T}_{\alpha\beta}(y) \right\rbrace\right\rangle \\
 +  \left\langle \mathbb{T}^* \left\lbrace \hat{T}_{\mu\nu}(x) \hat{T}_{\alpha\beta}(y) \right\rbrace\right\rangle -  \left\langle \tilde{\mathbb{T}}^* \left\lbrace \hat{T}_{\mu\nu}(x) \hat{T}_{\alpha\beta}(y) \right\rbrace\right\rangle .
\end {multline}
Here, $\lbrace,\rbrace$ is the anti-commutator and $\tilde{\mathbb{T}}$ is the reverse Weyl ordering computed by reversing the time ordering in Eq.~\eqref{Weyl_order}. Using Eq.~\eqref{complex}, we can decompose the Weyl ordered average into its real and imaginary parts
\begin {multline}
  \left\langle \mathbb{T}^* \left\lbrace \hat{T}_{\mu\nu}(x) \hat{T}_{\alpha\beta}(y) \right\rbrace\right\rangle =  \frac{1}{2} \left\langle \left\lbrace \hat{T}_{\mu\nu}(x),\hat{T}_{\alpha\beta}(y) \right\rbrace\right\rangle \\
 + i \textbf{Im} \left\langle \mathbb{T}^* \left\lbrace \hat{T}_{\mu\nu}(x) \hat{T}_{\alpha\beta}(y) \right\rbrace\right\rangle.
\end {multline}

Following Martin and Verdaguer ~\cite{Martin_Verdaguer_semiclassical_gravity} we define the (bi-tensor) real kernels. 
\begin {subequations}
\label{kernels_app}
\begin {align}
\hat{H}^S_{\mu\nu\alpha\beta} (x,y) &=   \textbf{Im} \left\langle \mathbb{T}^* \left\lbrace \hat{T}_{\mu\nu}(x) \hat{T}_{\alpha\beta}(y) \right\rbrace\right\rangle,\\
\hat{H}^A_{\mu\nu\alpha\beta} (x,y) &=   \frac{-i}{2} \left\langle  \left[ \hat{T}_{\mu\nu}(x), \hat{T}_{\alpha\beta}(y) \right]\right\rangle,\\
\label{response_app}
\hat{H}_{\mu\nu\alpha\beta} (x,y) &= \hat{H}^A_{\mu\nu\alpha\beta} (x,y) + \hat{H}^S_{\mu\nu\alpha\beta} (x,y), \\
 &= -i\theta(x^0 - y^0)  \left\langle  \left[ \hat{T}_{\mu\nu}(x), \hat{T}_{\alpha\beta}(y) \right]\right\rangle
 \end {align}
 \end {subequations}
\begin {subequations}
\begin {align}
\hat{N}_{\mu\nu\alpha\beta} (x,y) & = \frac{1}{2} \left\langle \left\lbrace \hat{t}_{\mu\nu}(x),\hat{t}_{\alpha\beta}(y) \right\rbrace\right\rangle, \\
\text{where} \quad \hat{t}_{\mu\nu}(x)  &= \hat{T}_{\mu\nu}(x) -  \left\langle\hat{T}_{\mu\nu}(x) \right\rangle,\\
\label{K_response_app}
\hat{K}_{\mu\nu\alpha\beta} (x,y) &= \frac{-4}{\sqrt{-g(x)} \sqrt{-g(y)}}
\left\langle \frac{\delta^2 S_{ph}[\hat{\phi}]}{ \delta g^{\mu\nu}(x) \delta g^{\alpha \beta}(y)} \right\rangle.
\end {align}
\end {subequations} 
The kernels $H^S$ and $H^A$ are the symmetric and anti-symmetric bi-tensor parts of $H$, i.e. 
\begin {equation}
H^{S,A}_{\mu\nu\alpha\beta}(x,y) = \pm H^{S,A}_{\alpha\beta\mu\nu}(y,x).
\end {equation}

With these, the second order variational derivative of the effective action on the forward branch reads
\begin {multline}
\label{plus_plus}
\frac{4 }{\sqrt{-g(x)} \sqrt{-g(y)}}\frac{\delta^2 \Gamma_{ph}}{ \delta (g^+)^{\mu\nu}(x) \delta (g^+)^{\alpha \beta}(y)}\bigg\rvert_{g^+ = g^-}\\ = i \hat{N}_{\mu\nu\alpha\beta} (x,y) -\hat{H}^S_{\mu\nu\alpha\beta} (x,y) - \hat{K}_{\mu\nu\alpha\beta} (x,y).
\end {multline}

Similarly,  the second order variation with respect to the other combinations of the forward and backward values of the metric are
\begin {multline}
\label{plus_minus}
\frac{4 }{\sqrt{-g(x)} \sqrt{-g(y)}}\frac{\delta^2 \Gamma_{ph}}{ \delta (g^+)^{\mu\nu}(x) \delta (g^-)^{\alpha \beta}(y)}\bigg\rvert_{g^+ = g^-} \\= -i \hat{N}_{\mu\nu\alpha\beta} (x,y) -\hat{H}^A_{\mu\nu\alpha\beta} (x,y),
\end {multline}
\begin {multline}
\label{minus_plus}
\frac{4 }{\sqrt{-g(x)} \sqrt{-g(y)}}\frac{\delta^2 \Gamma_{ph}}{ \delta (g^-)^{\mu\nu}(x) \delta (g^+)^{\alpha \beta}(y)}\bigg\rvert_{g^+ = g^-} \\= -i \hat{N}_{\mu\nu\alpha\beta} (x,y) +\hat{H}^A_{\mu\nu\alpha\beta} (x,y),
\end {multline}
\begin {multline}
\label{minus_minus}
\frac{4 }{\sqrt{-g(x)} \sqrt{-g(y)}}\frac{\delta^2 \Gamma_{ph}}{ \delta (g^-)^{\mu\nu}(x) \delta (g^-)^{\alpha \beta}(y)}\bigg\rvert_{g^+ = g^-} \\= i \hat{N}_{\mu\nu\alpha\beta} (x,y) +\hat{H}^S_{\mu\nu\alpha\beta} (x,y) + \hat{K}_{\mu\nu\alpha\beta} (x,y).
\end {multline}

The Eq.~\eqref{second_order} is a compact way to write the above second order variations Eqs.~\eqref{plus_plus}--~\eqref{minus_minus}.

\bibliographystyle{apsrev4-1}
\bibliography{references_sag}

\end{document}